\documentclass[conference]{IEEEtran}
\IEEEoverridecommandlockouts
\usepackage{cite}
\usepackage{amsmath,amssymb,amsfonts}
\usepackage{algorithmic}
\usepackage{graphicx}
\usepackage{textcomp}
\usepackage{xcolor,soul,framed}
\usepackage{comment}
\usepackage{makecell}
\usepackage{booktabs}
\usepackage[flushleft]{threeparttable}
\usepackage[linesnumbered,lined,boxed,commentsnumbered,ruled,longend,noend]{algorithm2e}
\usepackage{wrapfig}
\usepackage{bbm}

\usepackage{hyperref}
\hypersetup{hidelinks=true}

\SetCommentSty{mycommfont}
\SetKwInput{KwInput}{Input}                
\SetKwInput{KwOutput}{Output}              

\usepackage{fancyhdr,lipsum}

\fancypagestyle{mahmood}{%
   \fancyhf{} 
   
   \fancyhead[C]{You can use this material personally. Reprinting or republishing this material for the purpose of advertising or promotion, creating new collective works, reselling or redistributing to servers or lists, or using any copyrighted component in other works must adhere to IEEE policy. The DOI will be supplied as soon as it becomes available.}
}%

\makeatletter
\let\ps@IEEEtitlepagestyle\ps@mahmood
\makeatother

\begin{document}



\title{A Novel Multiple Access Scheme for Heterogeneous Wireless Communications using Symmetry-aware Continual Deep Reinforcement Learning}

\author{
    \IEEEauthorblockN{
        Hamidreza Mazandarani\textsuperscript{1}, Masoud Shokrnezhad\textsuperscript{2}, and Tarik Taleb\textsuperscript{3} \\
    }
    \IEEEauthorblockA{
       \textsuperscript{1} \textit{Ruhr University Bochum, Bochum, Germany; hr.mazandarani@ieee.org} \\
        \textsuperscript{2} \textit{ICTFicial Oy, Espoo, Finland; masoud.shokrnezhad@ictficial.com} \\
        \textsuperscript{3} \textit{Ruhr University Bochum, Bochum, Germany; tarik.taleb@rub.de} \\
    }
}

\maketitle

\begin{abstract}
The Metaverse holds the potential to revolutionize digital interactions through the establishment of a highly dynamic and immersive virtual realm over wireless communications systems, offering services such as massive twinning and telepresence. This landscape presents novel challenges, particularly efficient management of multiple access to the frequency spectrum, for which numerous adaptive Deep Reinforcement Learning (DRL) approaches have been explored. However, challenges persist in adapting agents to heterogeneous and non-stationary wireless environments. In this paper, we present a novel approach that leverages Continual Learning (CL) to enhance intelligent Medium Access Control (MAC) protocols, featuring an intelligent agent coexisting with legacy User Equipments (UEs) with varying numbers, protocols, and transmission profiles unknown to the agent for the sake of backward compatibility and privacy. We introduce an adaptive Double and Dueling Deep Q-Learning (D3QL)-based MAC protocol, enriched by a symmetry-aware CL mechanism, which maximizes intelligent agent throughput while ensuring fairness. Mathematical analysis validates the efficiency of our proposed scheme, showcasing superiority over conventional DRL-based techniques in terms of throughput, collision rate, and fairness, coupled with real-time responsiveness in highly dynamic scenarios.

\end{abstract}


\section{Introduction}
\IEEEPARstart{T}{he} Metaverse is an ever-evolving and transformative concept poised to establish a dynamic and immersive virtual realm, blurring the boundaries between the digital and physical domains and presenting an online environment so lifelike that it becomes virtually indistinguishable from reality \cite{tang_roadmap_2022, theodoropoulos2022cloud}. Expected to transcend the capabilities of the Internet, this paradigm harbors the potential to transform a multitude of service ecosystems across diverse facets of life, including immersive telepresence facilitated by virtual/extended reality \cite{taleb2022vr, taleb2022toward}, mobile augmented reality \cite{10304077}, and extensive twinning leading to intelligent industrialization \cite{10017413}. Consequently, it presents novel and unique challenges for the advancement of future wireless networks \cite{Metaverse_survey_2023}, which are already striving to deliver unprecedented levels of quality and capacity as well as remarkably low energy consumption \cite{giordani_toward_2020}. As the Metaverse encompasses diverse worlds, each offering a unique array of services, ensuring such consistent quality standards becomes paramount due to the ever-changing nature of this virtual realm. For instance, the traffic patterns and requirements within individual Metaverse User Equipments (UEs) and across multiple UEs, as they transition between services and virtual environments, may experience temporal shifts. UE mobility adds more complexity, leading to varying numbers of these UEs seeking to transmit data over the spectrum over time. Moreover, the conditions of the frequency channels themselves can change, influenced by a multitude of noise sources and environmental conditions.

Given this dynamic and ever-changing landscape, the realization of future wireless networks relies heavily on the implementation of adaptive multiple-access algorithms. In this challenge, dynamic UEs engage in continuous competition for access to either single or multiple frequency channels. In this context, where rapid decision-making within microseconds is imperative, conventional multiple-access techniques designed for stable conditions or with significant convergence times may fall short. Despite recent advances in network orchestration techniques \cite{shokrnezhad2022near, shokrnezhad2023scalable}, a new paradigm is needed. The concept of \textit{self-sustaining networking} emerges as a noteworthy approach for the 6th generation of wireless communication systems (6G) \cite{alwis_survey_2021}. This concept is enabled by adaptive Artificial Intelligence (Adaptive AI), transforming the traditional approach of \textit{once-in-a-lifetime } train models into a new paradigm of \textit{continuous re-training} with up-to-date data and evolving circumstances. \textcolor{black}{In other words, the learning problem will be viewed as continuous adaptation instead of seeking a fixed solution \cite{abel2024definition}}. Anticipated as a key catalyst for enabling the delivery of upcoming services, such as Metaverse applications \cite{alwis_survey_2021}, Adaptive AI has garnered significant attention in recent years. Notably, Gartner recognized it as one of the strategic technology trends for the year 2023 \cite{groombridge_gartner_nodate}.

In recent years, Deep Reinforcement Learning (DRL) has emerged as a potent tool for enabling adaptive multiple access to the frequency spectrum in heterogeneous environments. While ingenious techniques have been put forth, a critical challenge remains unattended: the ability to adapt agents to non-stationary environments. Given DRL's limitation in reusing previously acquired knowledge, accommodating each change could prove time-consuming, especially when contextual shifts occur frequently. Consequently, these aforementioned approaches prove unsuitable for Metaverse scenarios, where the environment exhibits remarkable dynamism. The highly fluid and ever-changing nature of the Metaverse demands more sophisticated solutions that can swiftly adapt to its volatile conditions. This paper fills in the gap in the current literature by presenting the following contributions:
\begin{itemize}
    \item We investigate the problem of multiple access in non-stationary, multi-channel, and heterogeneous wireless environments to maximize the intelligent agent throughput while maintaining fairness. Here, heterogeneity refers to the coexistence of multiple UEs with different protocols and transmission profiles unknown to each other, and non-stationarity is caused by intermittent changes in the set of active UEs and their transmission profiles.
    \item To solve the problem, an adaptive DRL-based Medium Access Control (MAC) protocol based on the Double and Dueling Deep Q-Learning (D3QL) algorithm empowered by a symmetry-aware Continual Learning (CL) mechanism is proposed. 
    \item To the best of our knowledge, our work stands as pioneering in harnessing CL to enhance DRL-based MAC protocols, tailored precisely to suit the distinctive attributes of the Metaverse. Despite the Metaverse's dynamic nature, its state space is finite (detailed in Section \ref{s_bck}-A). By defining contexts and leveraging symmetries, the proposed symmetry-aware CL mechanism exploits prior knowledge acquired throughout the agent's lifetime with a greater potential for backward transfer. We mathematically prove that using this mechanism, the state space can be reduced to unique contexts, making the agent more efficient and responsive to the dynamic, yet finite nature of the surrounding environment.
    \item We target to solve the problem considering that the behavior of incumbent UEs is beyond the agent’s control (due to considerations of backward compatibility and privacy preservation) and is governed by their hard-coded rules.
    \item Extensive simulations are conveyed to demonstrate that our proposed scheme can outperform conventional DRL-based techniques in terms of agent throughput, collision rate, and fairness constraints.
    \item Although our approach is designed for the Metaverse services, it can be adapted to other 6G dynamic services with similar strict Quality of Service (QoS) requirements. Furthermore, it introduces some ideas for dealing with non-stationary wireless networks.
\end{itemize}

\begin{figure*}[!t]
\centerline{\includegraphics[width=7.1in]{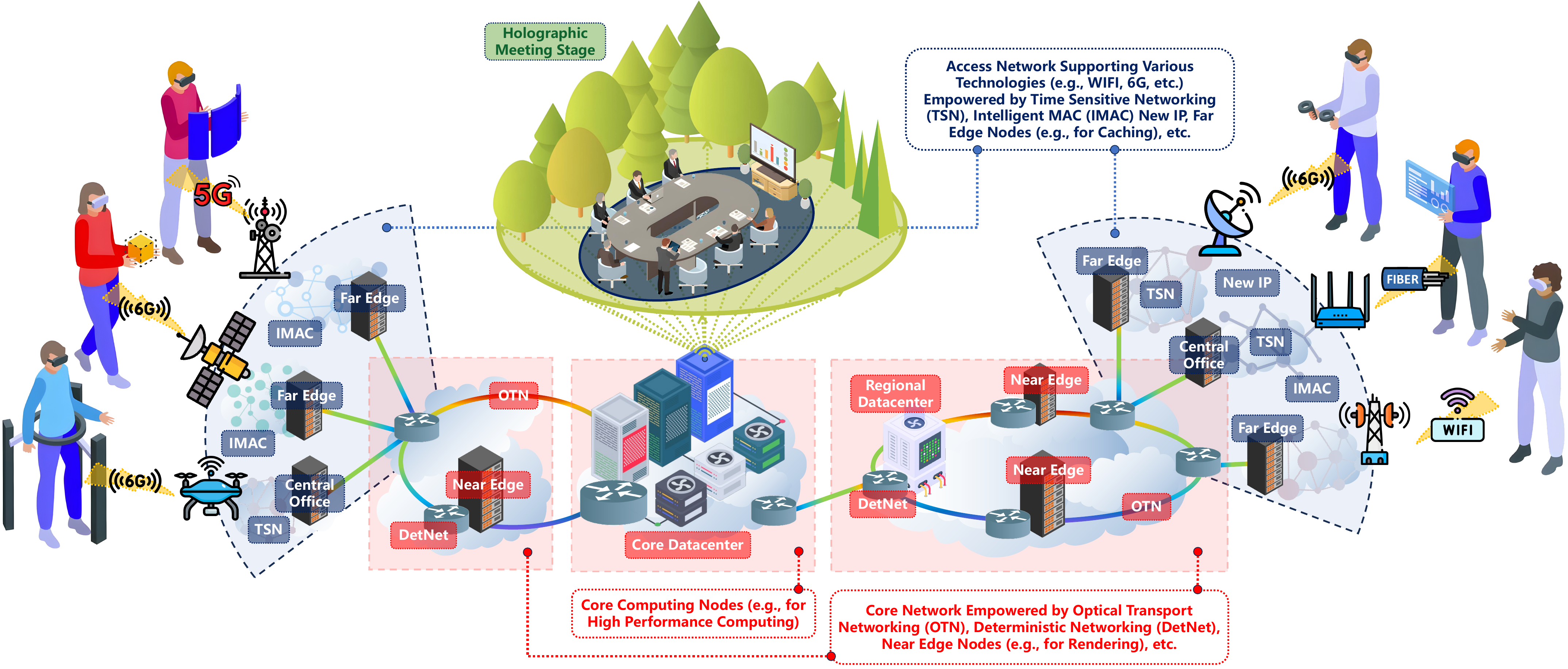}}
\caption{A lakeside holographic meeting room in the Metaverse enabled by a cloud-network integrated infrastructure powered by technologies including deterministic networking, time-sensitive networking, and intelligent medium access control, with end users connected via 6G, 5G, WiFi, and fiber connections.}
\label{fig01}
\end{figure*}

Notably, this paper is an extension of our previous work \cite{metacom_2023} with the same setup but with the following major enhancements:
\begin{itemize}
  \item A more heterogeneous environment is considered, comprised of UEs with various protocols, including more dynamic ones and those with complex and undeterminable patterns, such as Carrier-Sense Multiple Access (CSMA) or Channel Hopping (CH).
  \item The challenge is formulated as a novel optimization problem that considers variable packet length of UEs, resulting in explicit dependency between time slots.
  \item Fairness is considered in addition to agent throughput.
  \item The D3QL method is leveraged instead of the Double Deep Q-Learning (DDQL) algorithm.
  \item The CL mechanism undergoes a complete transformation by symmetry awareness, and its efficiency is mathematically investigated.
\end{itemize}

The remainder of this paper is organized as follows. Section \ref{s_bck} introduces the required background and the literature review. In Section \ref{s_prb}, the system model is introduced, and the problem is formulated and analyzed. The proposed strategy is then presented and investigated in Section \ref{s_app}. Finally, numerical results are illustrated and analyzed in Section \ref{s_sim}, followed by concluding remarks in Section \ref{s_con}.

\section{Background \& Related Work}\label{s_bck}
\subsection{Vision of 6G in the light of the Metaverse} \label{vision_of_6g}
As the long-term vision of digital transformation, the Metaverse is considered the inevitable fate of the Internet revolution and evolution \cite{tang_roadmap_2022, xu_full_2022}. Furthermore, the Metaverse is an amalgamation of various services and use-cases, rather than a single one \cite{yu2024attention}: 1) A fully immersive simulation of a plant production line, stretched out before the manager, projecting a seemingly endless expanse of conveyor belts and machines with the ability to control them in real time, or 2) a holographic meeting room on a lakeshore filled with the avatars of managers discussing product launch over augmented charts and pipelines are two ultimate illustrations of a Metaverse-enabled universe. Fig.~\ref{fig01} illustrates the holographic meeting scenario enabled through a cloud-network integrated infrastructure empowered by deterministic and time-sensitive networking \cite{shokrnezhad2022near, shokrnezhad2023double}. The 5G Public Private Partnership (5G-PPP) Architecture Working Group \cite{6G_Architecture}
identifies massive twinning and telepresence as inaugural Metaverse use cases \cite{nadir2021immersive, taleb2021extremely}. These services require microsecond-level latency, bounded jitter, multi-gigabit-level throughput, extremely high reliability, and extremely low energy consumption \cite{giordani_toward_2020}, with different levels of dependability on different factors for each service. 

In addition to the stringency of QoS requirements and their heterogeneity, another challenge to realizing the Metaverse is dynamicity\textcolor{black}{\cite{adil2024role}}. Following the virtual meeting example, new users can join and depart from the meeting dynamically, or their physical mobility at different velocities induces alterations in their points of connection to the system. Additionally, users have the flexibility to seamlessly shift between different roles, and the introduction of varied transmitters or alterations in meeting layouts occurs dynamically, thereby impacting users' transmission patterns. Despite these changes influencing the system's state, it is essential to note that the state does not proliferate indefinitely. For example, suppose there are $\mathcal{N}$ participants in a meeting, each of whom has adopted one of the $\Upsilon$ transmission profiles. By defining the system state as the set of active participants along with their transmission profiles, the total number of states would not be more than ${(\Upsilon + 1)^{\mathcal{N}}}$. Accordingly, some states will likely be experienced recurrently during a session. Hence, there is an opportunity to exploit prior knowledge in the future with the aim of improving QoS for users.

Despite the anticipated enhancements in capabilities, such as semantic awareness \cite{shokrnezhad2023semantic},  and integrated resource allocation \cite{shokrnezhad2024towards}, it remains imperative for 6G networks to align with the specific requirements and characteristics of services intended for delivery within the Metaverse \cite{tang_roadmap_2022}. Particularly in the presence of highly dynamic users, the network must demonstrate swift and dynamic adaptability, promptly adjusting its configuration, allocation, and resource utilization in response to immediate changes in system states. Spectrum sharing among a group of ever-fluctuating, all-primary UEs arises prominently in this context \cite{matinmikko2020spectrum}. Coordinating dynamic UEs accessing the same spectrum requires sophisticated interference management and robust contention resolution mechanisms to ensure their required QoS. One prospective strategy to meet this challenge involves the incorporation of self-x capabilities (such as self-management, self-planning, self-organization, self-optimization, self-healing, and self-protection), resulting in the formation of a self-sustaining 6G network. A true self-sustaining 6G network would be built upon intelligent learning approaches with the capability of exploiting prior knowledge while adapting to novel occurrences.

\begin{table*}[!hbt]
\caption{Selected Papers Comparison.}
\centering
\begin{tabular}{|c|c|c|c|c|c|c|}

\hline

\textbf{Reference} & \textbf{Objective(s)} & \textbf{Incumbent UEs/Networks} & \makecell{\textbf{Multi-} \\ \textbf{channel}} & \makecell{\textbf{Variable} \\ \textbf{Packet Length}} & \makecell{\textbf{Time-varying} \\ \textbf{Environment}} & \makecell{\textbf{\color{black} Symmetry} \\ \textbf{\color{black} awareness}} \\

\hline

DLMA \cite{yu_deep-reinforcement_2019} & \makecell{Sum Throughput + \\ Proportional Fairness} & TDMA + Different Versions of ALOHA & - & - & - & - \\

\hline

CS-DLMA \cite{yu_non-uniform_2021} & \makecell{Sum Throughput + \\ Proportional Fairness} & TDMA + CSMA Networks & - & \checkmark & - & - \\

\hline

SOMAC \cite{gomes2020automatic} & \makecell{Sum Throughput + \\ Delay} & TDMA + CSMA UEs & - & - & \checkmark & - \\

\hline

Chen \textit{et al.} \cite{chen2022dueling} & Intelligent UE Throughput & \makecell{Various Networks Containing \\ TDMA, CH, \\ and Stochastic UEs} & \checkmark & - & \checkmark & - \\

\hline

MC-DLMA \cite{ye2021multi} & Sum Throughput & \makecell{Various Networks Containing \\ TDMA, Different Versions of \\ ALOHA and Other Intelligent Agents} & \checkmark & - & \checkmark & - \\

\hline

\color{black} HD-RL \cite{ni2024dynamic} & \makecell{\color{black} Throughput \\ \color{black} and fairness} & \makecell{\color{black} TDMA and \\ \color{black} q-ALOHA UEs} & - & - & - & -\\

\hline

\color{black} CHMA \cite{han2024multiple} & \makecell{\color{black} Average Throughput \\ \color{black} and Fairness} & \makecell{\color{black} TDMA and \\ \color{black} ALOHA UEs} & \color{black} \checkmark & \color{black} - & \color{black} \checkmark & - \\

\hline

CL-DDQL \cite{metacom_2023} & \makecell{Sum Throughput} & \makecell{TDMA UEs} & \checkmark & \checkmark & \checkmark & - \\

\hline

\makecell{Our Approach \\ (CL-D3QL)} & \makecell{Agent Throughput \\ with Constrained Fairness} & \makecell{Various TDMA, CSMA and \\ CH UEs} & \checkmark & \checkmark & \checkmark & \checkmark \\

\hline
\end{tabular}
\label{tab_paper_comparison}
\begin{tablenotes}
\footnotesize
\item \cite{metacom_2023} is our previous work.
\end{tablenotes}
\end{table*}

\textcolor{black}{In the same vein, MAC protocols should be tailored to mostly distinct demands of the Metaverse, particularly in supporting dynamicity due to frequent context transitions, and inter-operability among various subsystems \cite{xu_full_2022, adil2024role}. Therefore, an ideal MAC for Metaverse should:
\begin{itemize}
    \item Support multi-channel configuration for enhanced multi-modal data transmission rates;
    \item Coexist with multiple devices using unknown protocols;
    \item Quickly adapt to ever-ending environmental context transitions, including fluctuations in the number of active users \cite{abel2024definition}.
\end{itemize}
In the following, we briefly outline efforts toward an ideal MAC protocol and will focus on leveraging AI for this in the next subsection.
The emerging field of protocol learning is gaining momentum, categorized by Park \textit{et al.} \cite{park2024towards} into three levels: task-oriented neural protocols, symbolic protocols, and semantic protocols that leverage Large Language Models (LLMs). Additionally, socially aware human-centered resource allocation introduced in \cite{yu2023socially} can further enhance MAC performance in the Metaverse ecosystem. Moreover, integrating Metaverse application information into the MAC process as well as predicting upcoming services is another way to improve the scheme \cite{yang2021brainiot}. However, such mechanisms, particularly leveraging LLMs, are not always feasible in wireless networks with non-intelligent legacy devices.}

\subsection{AI-based Adaptive MAC Protocols}
In recent years, enabling networks with self-sustaining capabilities has been extensively studied in the literature to investigate the problem of adaptive multiple access to the frequency spectrum \cite{alwarafy2022frontiers}. For instance, Yu \textit{et al.} \cite{yu_deep-reinforcement_2019} adopted DRL to design a MAC protocol without assuming the protocol of other coexisting UEs. They considered a heterogeneous environment consisting of their designed agent called DLMA, with a few Time-Division Multiple Access (TDMA) and ALOHA UEs competing over a slotted uplink channel. The same authors extended their work to non-uniform scenarios, in which channel sensing requires a single time slot but information packet transmission requires multiple time slots \cite{yu_non-uniform_2021}. Following the idea of utilizing DRL for various scenarios of adaptive multiple access problems, Jadoon \textit{et al.} \cite{jadoon2022deep} utilized DRL to optimize both throughput and packet age. Their research is compatible with machine-type communications on the assumption that the UEs are not saturated. Doshi \textit{et al.} \cite{doshi2021deep} formulated the coexistence of multiple base stations over a shared channel, optimizing the signal-to-interference-plus-noise ratio of UEs. Besides, Guo \textit{et al.} \cite{guo_multi-agent_2022} developed a solution for multi-agent scenarios to support delay-sensitive requests. Gomes \textit{et al.} \cite{gomes2020automatic} proposed SOMAC, an RL-based algorithm that switches between TDMA and CSMA protocols according to the network situation. Chang \textit{et al.} \cite{chang2023federated} incorporate federated learning and multi-agent DRL to design a collaborative distributed spectrum access strategy. Authors of \cite{9241414} propose a multi-dimensional intelligent multiple access considering disparate resource constraints among heterogeneous equipment. \textcolor{black}{Ni \textit{et al.} \cite{ni2024dynamic} present HD-RL, a dynamic wireless channel-sharing solution that leverages lightweight hyperdimensional computing to enhance spectrum utilization and channel capacity in wireless networks. Han \textit{et al.} \cite{han2024multiple} introduce CHMA protocol, which integrates curriculum learning (i.e., a learning technique that gradually increases the difficulty level of tasks) and reinforcement learning to enhance the throughput and fairness of the system.}

Additionally, there are research proposals in the existing literature that extend AI-based MACs to multi-channel scenarios. Because every UE is only aware of the channel on which it resides, this scenario has an inherent difficulty that can be described as partial observability \cite{chen2022dueling}. As an instance, Chen \textit{et al.} \cite{chen2022dueling} proposed a dueling deep recurrent Q-Network to solve the PSPACE-hard problem of multiple access in the presence of various co-existing networks. Ye \textit{et al.} \cite{ye2021multi} investigated the same problem and proposed the MC-DLMA protocol that outperforms the random access policy, the Whittle index policy, and the original DQN. The literature on DRL-based adaptive MAC is becoming a rich research area, expanding in many directions including the number of intelligent agents (i.e., multi-agent RL), number of channels, compatibility with standardized mechanisms, customization for special scenarios such as the Internet of Things (IoT), etc. In this vein, the interested reader may find \cite{zheng2023survey} and \cite{abbasi2021deep} insightful. In Table \ref{tab_paper_comparison}, we provide a comparison among selected MAC protocols for heterogeneous environments, wherein the intelligent agent coexists alongside multiple UE types using unknown MAC protocols.

Even though the heterogeneity problem has been addressed in existing literature, the non-stationary nature of Metaverse services has remained mostly untouched. \textcolor{black}{This paper fills in the gap by introducing a novel approach that takes into account the repetition of states over time, providing deeper insights into temporal dynamics and the patterns that arise from these repetitions.} Similar to Yu \textit{et al.} \cite{yu_non-uniform_2021}, we assume UEs with variable-length packets as this is more practical than fixed-length packets \cite{yu_deep-reinforcement_2019}. Unlike Yu \textit{et al.} \cite{yu_non-uniform_2021}, however, our approach takes multiple channels into account, making it even more applicable in high-bandwidth Metaverse environments. \textcolor{black}{Furthermore, we leveraged intrinsic symmetries in the multiple-channel setup to develop more efficient context management, a previously unexplored approach in the existing literature. Finally, in line with the literature, we assume a saturated traffic pattern where users continuously transmit information streams, with task management occurring at the upper layer. This assumption also aligns with Metaverse applications like holographic meetings (illustrated in Fig.~\ref{fig01}), where users utilize UDP-like transmission protocols.}


\section{Problem Definition}\label{s_prb}
\subsection{System Model}

\begin{figure}[!t]
\centerline{\includegraphics[width=3.2in]{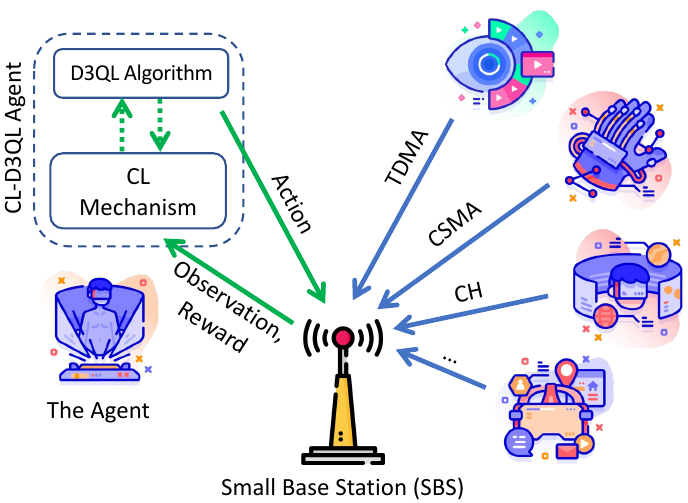}}
\caption{The system model.}
\label{fig_sm}
\end{figure}

We consider a single small cell covered by a Small Base Station (SBS) with $\mathcal{N}$ heterogeneous UEs each uniquely labeled as $u_{i}, i \in \mathbb{N} = \{0, …, \mathcal{N}\}$ (see Fig. \ref{fig_sm}). Except for one (i.e., the CL-D3QL agent, or simply the \textit{agent}, denoted with ${u}_{0}$), all UEs are assigned a channel for each time slot among $\mathcal{C}$ time-slotted channels $\in \mathbb{C} = \{0, …, \mathcal{C}\}$ and designed to occupy that channel by transmitting their packets using either Time-Division Multiple Access (TDMA), Carrier-Sense Multiple Access (CSMA), or Channel Hopping (CH) protocols. As an illustrative example, we consider the Metaverse scenario described in section \ref{vision_of_6g} wherein various services with different QoS requirements and multiple access protocols coexist. In this ecosystem, TDMA protocol can be used within different services to integrate physical and virtual worlds through digital twinning \cite{xu_full_2022}. For instance, a headset may send visual recordings to its control center every millisecond. Conversely, CSMA and CH are more complex, yet non-intelligent protocols deployed to offer more advanced services. For the sake of backward compatibility and privacy preservation, the internal functionality and microstates of incumbent protocols (i.e., TDMA, CSMA, and CH) are beyond the scope of agent observation and control.

The environment is non-stationary due to the fluctuating number of active incumbent UEs, in addition to the transmission profiles they may adopt. Frequent changes in the number of active users or transmission profiles, which is common in the Metaverse, can be attributed to various factors, including users' mobility, role, layout, and transceivers, as mentioned in the previous sections. As an example in the headset scenario, if the user changes the role from an actor to a viewer, the UE may proceed to the inactive mode to only provide vital functionalities; thus, data may be transmitted every ten milliseconds. A \textit{context} is defined as a collection of active UEs with unique identifiers on specific channels. For instance, if we have two UEs $\{0, 1\}$ and two channels $\{1, 2\}$, UE $0$ on channel $1$ and UE $1$ on channel $2$ would constitute a simple context, whereas having only UE $0$ active in the network sending over channel 1 would be a different context. Context transitions occur when a UE enters or leaves a channel or changes its transmission profile\footnote{Changing the transmission profile will create a new identifier for UEs (for the sake of simplicity).}. In this paper, it is posited that changes in expected throughput are the primary driver of changes in transmission profiles.

It is assumed that the agent is informed of the arrivals and departures of other UEs on each channel along with their expected throughputs via the SBS\footnote{ While the SBS has complete knowledge of the past, it has no knowledge of the future, including which users will transmit packets with which lengths and transmission profiles in upcoming time slots.}. However, the agent is unaware of the transmission protocols and packet lengths. Note that we assume UEs with variable-length packets. The agent's transmissions are independent of the SBS to avoid unnecessary signaling overhead in scheduling grant decoding. However, it relies on the SBS's ACK signals issued at the end of each packet transmission (or channel sensing) to indicate successful transmission (or channel idleness). The transmission of control messages is assumed to occur over a separate, collision-free channel.

\subsection{Problem Formulation}

In this subsection, we introduce a Mixed-Integer Non-Linear Programming (MINLP) formulation to define the problem. The subsequent subsection will provide additional details on the variables, constraints, and objective function of the problem.

\subsubsection{Decision and Support Variables}

The primary integer decision variable in our problem is $\mathbb{R} = [{r}_{c}^{t}]_{\mathcal{C} \times \mathcal{T}}$, where each element represents the size of the packet that the agent (i.e., $u_{0}$) begins to transmit on channel $c$ at the start of time slot $t$. Additionally, $\mathbb{Z} = [{z}_{c}^{t}]_{\mathcal{C} \times \mathcal{T}}$ denotes an auxiliary integer variable indicating the remaining time slots for the ongoing transmission. The variable $\mathbb{M} = [{m}_{c}^{t}]_{\mathcal{C} \times \mathcal{T}}$ serves as another auxiliary variable, set to one whenever ${z}_{c}^{t}$ is positive. The following constraints transform the equation ${m}_{c}^{t} = \mathbbm{1}({z}_{c}^{t} > 0)$ into a linear form:
\begin{align}\label{z_m_constraints}
    &{m}_{c}^{t} \leq {z}_{c}^{t}, \\
    &{m}_{c}^{t} \cdot R_{max} \geq {z}_{c}^{t}, \notag
\end{align}
where $R_{max}$ is the maximum allowed packet size. Table \ref{tab_symbols} demonstrates all variables and constants, where symbols marked with $\hat{}$ sign are constants that are determined outside of the problem scope.

\subsubsection{Temporal Interdependency Constraints}

Now, we need to establish interdependency over time. First, we must ensure that $z$ decreases by one with each passing time slot for ongoing transmissions. This requirement is enforced by the following constraint.
\begin{align}\label{z_constraints}
{z}_{c}^{t} = {z}_{c}^{t - 1} - 1 \quad \text{if } \ \big( ( {m}_{c}^{t} = 1 ) \ \land \ ( {d}_{c}^{t} = 0 ) \big)
\end{align}
This constraint can be expressed in its linear equivalent forms as follows:
\begin{align}\label{z_constraints_linear}
    &{z}_{c}^{t} \leq ( {z}_{c}^{t - 1} - 1 ) + \left( R_{max} \cdot \big( (1 - {m}_{c}^{t} ) + {d}_{c}^{t} \big) \right), \\
    &( {z}_{c}^{t - 1} - 1 ) \leq {z}_{c}^{t} + \left( R_{max} \cdot \big( (1 - {m}_{c}^{t} ) + {d}_{c}^{t} \big) \right), \notag
\end{align}
where $\mathbb{D} = [{d}_{c}^{t}]_{\mathcal{C} \times \mathcal{T}}$ is a binary variable used to activate decision making over specific time slots.

Second, it must be ensured that a new transmission is not initiated until the ongoing one is completed. To achieve this, we establish a set of constraints based on the premise that whenever the agent initiates the transmission of a packet with length $k$ at time slot $t$, the activation variables for the subsequent $k-1$ time slots must be zero (i.e., ${{d}_{c}^{\tau}} = 0 \ \forall \tau \in [t+1, t+k-1]$):
\begin{align}\label{r_d_constraints}
    {r}_{c}^{t} = k \to  \sum_{\tau=t+1}^{t+k-1}{{d}_{c}^{\tau}} = 0 \quad \forall k \in [1, R_{max}]
\end{align}
This can be expressed in its linear equivalent form as follows (where $K_{1}$ and $K_{2}$ are auxiliary binary variables, both equal to one when ${r}_{c}^{t} = k$):
\begin{align}\label{r_d_constraints_linear}
    &( {r}_{c}^{t} - k ) + 1 \leq R_{max} \cdot K_{1} \notag \\
    &( k - {r}_{c}^{t} ) + 1 \leq R_{max} \cdot K_{2} \notag \\ 
    &\sum_{\tau=t+1}^{t+k-1}{{d}_{c}^{\tau}} \leq R_{max} \cdot (2 - ( K_{1} + K_{2}))
\end{align}

\subsubsection{Packet Length Constraints}

\begin{figure}[t!]\centering
\includegraphics[width=3.0in]{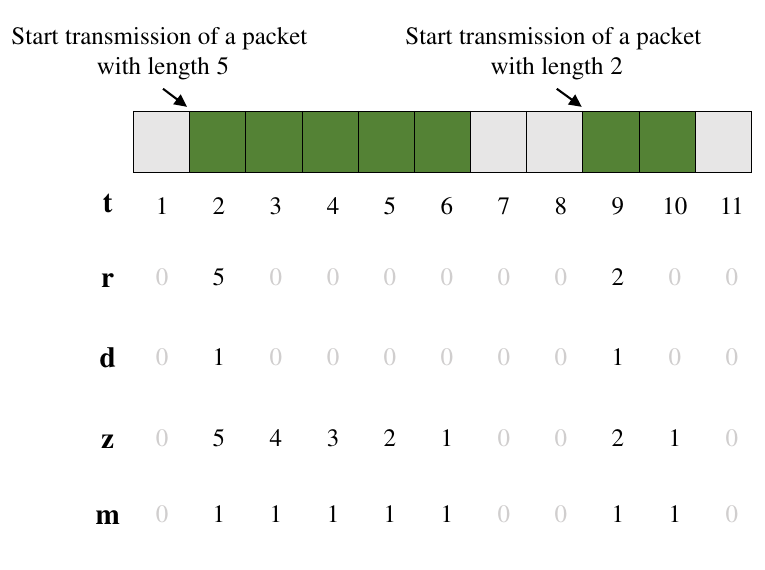}
    \vspace{-5pt}
    \caption{A scenario in which the agent initiates the transmission of packets with lengths $5$ and $2$ at time slots $t = 2$ and $t = 9$, respectively. The values of the decision variable $r$, along with the support variables $d$, $z$, and $m$, can be compared accordingly. \textcolor{black}{At the start of transmissions, both $r$ and $z$ equal the packet length, but $z$ decrements each time slot to enable the calculation of the per time slot transmission indicator $m$.}}
    \vspace{0pt}
    \label{variables}
\end{figure}

Next, we establish the relationship between the packet length variable $r$ and the support variables $z$ and $d$ using \eqref{r_constraints}, where $r$ is equal to $z$ when transmission starts:
\begin{align}\label{r_constraints}
{r}_{c}^{t} = \begin{cases} {z}_{c}^{t} & \text{if } \ {d}_{c}^{t} = 1 \\
0 & \text{if } \ {d}_{c}^{t} = 0 \end{cases}
\end{align}
We then transform \eqref{r_constraints} into a linear form, similar to previous constraints:
\begin{align}\label{r_constraints_linear}
    &{r}_{c}^{t} \leq {z}_{c}^{t}  + \left( R_{max} \cdot (1 - {d}_{c}^{t} ) \right) \\
    & {z}_{c}^{t} \leq {r}_{c}^{t}  + \left( R_{max} \cdot (1 - {d}_{c}^{t} ) \right) \notag
    \\
    & {r}_{c}^{t} \leq R_{max} \cdot {d}_{c}^{t} \notag
\end{align}
A simple example, as illustrated in Fig. \ref{variables}, demonstrates the values of $r$ and the other auxiliary variables. It is important to note that in a single time-slot scenario, the variables $r$, $z$, $m$, and $d$ are equal and binary.

\subsubsection{Physical Constraints}

Constraints \eqref{max_channel_constraints} and \eqref{max_user_constraints} ensure that the agent can transmit on at most one channel during each time slot, and the agent can transmit on a channel not occupied by an incumbent UE (as specified by $\hat{m}_{i, c}^{t}$). \textcolor{black}{These constraints enhance system scalability by narrowing the search space for potential transmissions.}
\begin{equation} \label{max_channel_constraints}
    \sum_{c \in \mathbb{C}} {m}_{c}^{t} \leq 1 \quad \forall t \in [0, \mathcal{T}]
\end{equation}
\begin{equation} \label{max_user_constraints}
    {m}_{c}^{t} \leq 1 - \mathbbm{1} \big( \sum_{i \in \mathbb{N} - \{ 0 \} } \hat{m}_{i, c}^{t} > 0 \big) \quad \forall c \in \mathcal{C}, \forall t \in [0, \mathcal{T}]
\end{equation}

\subsubsection{Normalized Throughputs Constraints}

The expected throughput of incumbent UEs is set according to their hard-coded rules, as shown in Table \ref{tab_expected_throughputs}. TDMA and CSMA nodes transmit on a single channel, while CH nodes split their traffic equally across the channels. The expected throughput of the intelligent agent is set to $1$, as it is expected to reach the throughput of $1$ in the absence of other UEs.

Eq. \eqref{normalized_throughputs} restricts the normalized throughput of the agent (i.e., its actual throughput divided by target (fair) throughput) to one for all time slots, prohibiting it from aggressive behavior. Target throughputs are calculated by a water-filling algorithm $f(.)$, guaranteeing that each UE receives a minimum of available channel share and its expected throughput, as detailed in Algorithm \ref{water_filling}. This function receives $\widehat{\boldsymbol{\mathcal{X}}}^{t}=\{\widehat{x}^{t}_{0, 0}, \widehat{x}^{t}_{0, 1}, ..., \widehat{x}^{t}_{\mathcal{N, C}}\}$, and returns a vector indicating the \textit{target} throughput of all UEs for all channels\footnote{As an example, assume that $\widehat{\boldsymbol{\mathcal{X}}}^{t}=[0.9,0.5,0.1]$ in a single-channel network. The water-filling algorithm returns $[0.45,0.45,0.1]$, with the justification that the third UE does not need more than 10\%, and the rest is divided equally between the first and second UEs. Note that these calculations are extensible to situations where different UEs have different priorities by assigning non-equal weights in the water-filling algorithm.}. \textcolor{black}{Remarkably, since ${\chi}_{c}^{t}$ in the denominator of \eqref{normalized_throughputs} is derived from the heuristic Water-filling algorithm, it does not introduce additional complexity to the problem to be defined in the next section.}

\begin{align} \label{normalized_throughputs}
{x}_{c}^{t} &= \frac{1}{h} \ \cdot \sum_{\tau \in [t - h, t]}{r_{c}^{\tau}} \notag \\
{\chi}_{c}^{t} &= f(\widehat{\boldsymbol{\mathcal{\chi}}}^{t})[0, c] \notag \\
\frac{{x}_{c}^{t}}{{\chi}_{c}^{t}} &\leq 1 \begin{array}{l} \forall c \in \mathbb{C}, \forall t \in [h, \mathcal{T}] \end{array}
\end{align}

\begingroup

\begin{table}[t!]
\caption{Expected Throughputs for different protocols}
\vspace{-1em}
\setlength\tabcolsep{2.5pt}
\renewcommand{\arraystretch}{2}
\begin{center}
\begin{tabular}{|c|c|}
\hline
\textbf{Protocol} & \textbf{Expected Throughputs} \\
\hline
TDMA & $\dfrac{\text{packet length}}{\text{duty cycle}}$ on single channel \\ \hline
CSMA & $\dfrac{\text{packet length}}{\text{packet length} + \dfrac{\text{window size}}{2}}$ on single channel \\ \hline
CH & $\dfrac{1}{\mathcal{C}}$ on all channels \\ \hline
CL-D3QL & $1$ on all channels \\ \hline
\end{tabular}
\label{tab_expected_throughputs}
\end{center}
\end{table}

\endgroup

\subsubsection{The Problem}

The problem is to maximize the intelligent agent throughput over its lifetime, denoted by $\mathcal{T}$ time slots, taking into account the channels and transmission profiles of the incumbent UEs without degrading their actual throughput, as formulated in \eqref{problem2}.

\begin{align}\label{problem2}
    &
    \max_{r} \sum_{h \leq t \leq \mathcal{T}} \sum_{c \in \mathbb{C}}  {x}_{c}^{t} \quad \mbox{s.t:} \\
    &
    \small\text{Variables' interdependencies acc. to } \eqref{z_m_constraints}, \eqref{z_constraints_linear}, \eqref{r_d_constraints_linear}, \eqref{r_constraints_linear}, \notag \\
    &
    \small\text{Physical channel constraints acc. to } \eqref{max_channel_constraints}, \eqref{max_user_constraints}, \notag \\
    &
    \small\text{Fair behavior constraint acc. to } \eqref{normalized_throughputs} \notag
\end{align}

\begingroup

\begin{table}[t!]
\caption{Notations of Symbols Used in the Problem.}
\vspace{-1em}
\setlength\tabcolsep{2.5pt}
\renewcommand{\arraystretch}{2}
\begin{center}
\begin{tabular}{|c|c|}
\hline
\textbf{Symbol} & \textbf{Description} \\
\hline
${r}_{c}^{t}$ & \makecell{Size of the packet the agent starts transmitting\\ on channel $c$ at the beginning of time slot $t$} \\ \hline
${z}_{c}^{t}$ & \makecell{Number of time slots remaining until\\ the current transmission on channel $c$ ends} \\ \hline
${m}_{c}^{t}$ & \makecell{An indicator for transmission of the agent\\ on channel $c$ in time slot $t$} \\ \hline
$\hat{m}_{i, c}^{t}$ & \makecell{A constant indicator for transmission of UE $i \in \mathbb{N} - \{ 0 \}$ \\ on channel $c$ in time slot $t$} \\ \hline
${d}_{c}^{t}$ & \makecell{An indicator for decision-making of the agent\\ on channel $c$ in time slot $t$} \\ \hline
${x}_{c}^{t}$ & Actual throughput of the agent on channel $c$ \\ \hline
${\chi}_{c}^{t}$ & Target (fair) throughput of the agent on channel $c$ \\ \hline
$\widehat{\chi}_{i, c}^{t}$ & Expected throughput of UE $i$ on channel $c$ \\ \hline
\end{tabular}
\label{tab_symbols}
\end{center}
\end{table}

\endgroup

\begin{algorithm}[t!]\label{water_filling}
\caption{Water-Filling}
\KwInput{$\widehat{\boldsymbol{\mathcal{X}}}$}
\KwResult{$\boldsymbol{\mathcal{X}}$}
$x_{i, c} \gets 0 \ \forall i \in \mathbb{N}, \forall c \in \mathbb{C}, \ \epsilon \gets 0.01$\\

\For{$c \in \mathbb{C}$}{
$r \gets 1$ \\
\While{$r > 0$} {
\For{$i \in \mathbb{N}$}
{
    \If{$x_{i, c} < \widehat{x}_{i, c}$}
    {
    $x_{i, c} = x_{i, c} + \epsilon$ \\
    $r = r - \epsilon$ \\}
}
}
}

\end{algorithm}

\subsection{Complexity Analysis}
Assuming that all the information of all time slots is known beforehand, or equivalently, all incumbent UEs have known deterministic transmission patterns, problem \eqref{problem2} could be solved with linear programming methods. With this assumption, the agent would know in which time slots and on which channels it must transmit data to maximize its throughput while satisfying the fairness constraint (Eq. \eqref{normalized_throughputs}). However, our problem belongs to the class of Partial Observable Markov Decision Problems (POMDP). Partial Observability is due to the following reasons. First, the agent is only aware of the channel it is sensing or transmitting a packet on. Secondly, the result of each multi-slot transmission is only known at the end. Thirdly, the system state is known till the time slot we are already in. The agent does not know what UEs will arrive in the coming time slots and what transmission profiles they will take. Lastly, the reaction of some other UEs is excessively complex to determine in an unknown environment. In particular, CSMA UEs adjust their window sizes randomly after a collision. Without this information, we cannot optimally solve problem \eqref{problem2}. Nonetheless, since the problem complexity has a linear relationship with time; the number of incumbent users; and the number of channels, scalable approaches are preferred.

\section{Solution Approach}\label{s_app}
\subsection{Learning Mechanism}
To handle the high level of self-sustaining required to address the heterogeneity and dynamism of the Metaverse environment \cite{xu_full_2022} and solve problem \eqref{problem2} in practical scenarios, RL is the learning mechanism adopted in this paper. An agent in RL learns through trial and error how to optimize a given decision-making problem (e.g., multiple access in wireless networks). The designer of the system specifies the reward function regarding the predefined design goals, and by learning and following the optimal strategy, the agent will maximize cumulative discounted rewards starting from any initial state. Q-learning is probably the most recognized among the different algorithms introduced for model-free RL problems \cite{watkins_q-learning_1992}. Each state-action pair is assigned a numeric value in Q-Learning, known as the Q value, and this value is gradually updated by the following equation, which is the weighted average of the old value and the new information, that is
\begin{equation}\label{eq_bellman}
Q(s^t, a^t)\; \mathrel{{+}{=}} \; \sigma[Y^t_{QL} - Q(s^t, a^t)],
\end{equation}
where $s^t$ and $a^t$ are the agent's state and action at time slot $t$ respectively, $\sigma$ is a scalar step size, and $Y^t_{QL}$ is the target, defined by
\begin{equation}\label{eq_target}
Y^t_{QL} = {\rho}^{t+1} + \gamma \; \text{max}_{a \in \boldsymbol{\mathcal{A}}} Q(s^{t+1}, a),
\end{equation}
where ${\rho}^{t+1}$ is the earned reward\footnote{In the literature, the reward is typically represented by $r$, but here we use the Greek symbol $\rho$ to avoid confusion with variable $r$ used for packet size.} at time slot $t+1$, $\gamma \in [0,1]$ is a discount factor that balances the importance of immediate and future rewards, and $\boldsymbol{\mathcal{A}}$ is the set of actions.

Since the problem defined in (\ref{problem2}) is impractical to discover all possible combinations of states and actions and learn all state-action values, DQL is a ground-breaking improvement to approximate them, in which a Deep Neural Network (DNN) is used as an approximator for Q values \cite{mnih_human-level_2015}. In DQL, the state is provided as the input, and the DNN-based $Q$ function of all possible actions, denoted by $Q(s, .; \boldsymbol{\mathcal{W}})$, is generated as the output, where $\boldsymbol{\mathcal{W}}$ is the set of DNN parameters. The target of DQL is
\begin{equation}\label{eq_DQL_target}
Y^t_{DQL}  = {\rho}^{t+1} + \gamma \; \text{max}_{a \in \boldsymbol{\mathcal{A}}} Q(s^{t+1}, a; \boldsymbol{{\mathcal{W}}^t}^-),
\end{equation}
and the update function of $\boldsymbol{\mathcal{W}}$ is
\begin{align}
\label{eq_DQL_bellman}
\boldsymbol{\mathcal{W}}^{t+1} = & \; \boldsymbol{\mathcal{W}}^{t} \nonumber \\
& + \sigma[Y^t_{DQL} - Q(s^t, a^t; \boldsymbol{\mathcal{W}}^t)]\nabla_{\boldsymbol{\mathcal{W}}^t} \nonumber \\
& \cdot Q(s^t, a^t; \boldsymbol{\mathcal{W}}^t).
\end{align}
To improve the efficiency and stability of DQL, the observed transitions are stored in a memory bank known as the experience memory, and the neural network is updated by randomly sampling from this pool \cite{mnih_human-level_2015}.

To further increase DQL's efficiency, we employ the idea proposed by Hasselt \textit{et al.} \cite{hasselt_deep_2016}, resulting in Double DQL. In both standard Q-Learning and DQL, the max operator is used to select and evaluate actions with the same values (or the same Q). As a result, optimistic value estimates are more likely to be chosen, due to the greater likelihood of selection of the overestimated values. Double DQL implements decoupled selection and evaluation processes. The following is the definition of the target in Double DQL.
\begin{equation}\label{eq_DDQL_target}
Y^t_{DDQL}  = {\rho}^{t+1} + \gamma \; \widehat{Q}(s^{t+1}, a'; \boldsymbol{{\mathcal{W}}^t}^-),
\end{equation}
where $a' = \text{argmax}_{a \in \boldsymbol{\mathcal{A}}} Q(s^{t+1}, a; \boldsymbol{\mathcal{W}}^t)$, and the update function is resulted by replacing $Y^t_{DQL}$ with $Y^t_{DDQL}$ in \eqref{eq_DQL_bellman}. In this model, $\boldsymbol{\mathcal{W}}$ represents the set of weights for the main (or evaluation) $Q$ and is updated in each step, whereas $\boldsymbol{\mathcal{W}}^-$ is for the target $\widehat{Q}$ and is replaced with the weights of the main network every $\hat{t} \gg 0$ steps. In other words, $\widehat{Q}$ remains a periodic copy of $Q$.

\begin{figure}[t!]\centering
\includegraphics[width=3.5in]{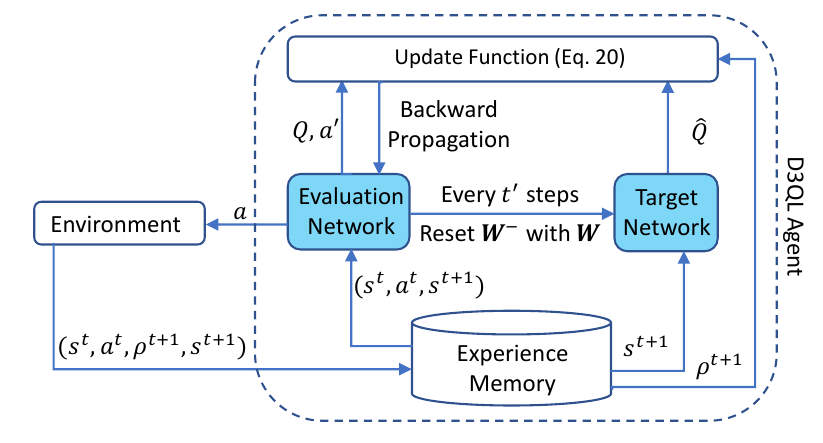}
  \caption{The D3QL agent.}
  \label{fig_d3ql}
\end{figure}

In addition, we enhance Double DQL by integrating it with the dueling idea proposed by Wang \textit{et al.} \cite{wang2016dueling}. In contrast to the Double DQL, where Q values are directly approximated by DNNs, this method first calculates two separate estimators for state values and action advantages, denoted by $\mathcal{V}$ and $\mathcal{A}$ respectively, and then determines Q values based on these estimators as \eqref{eq_dueling}, where $\boldsymbol{\mathcal{A}}$ is the action space. 
\begin{align}
\label{eq_dueling}
Q(s^t, a^t; \boldsymbol{\mathcal{W}}^t) = & \; \mathcal{V}(s^t; \boldsymbol{\mathcal{W}}^t) \nonumber \\
& + \Bigg( \mathcal{A}(s^t, a^t; \boldsymbol{\mathcal{W}}^t) \nonumber \\
& \qquad - \frac{1}{\left| \boldsymbol{\mathcal{A}} \right|} \sum_{a^{\prime} \in \boldsymbol{\mathcal{A}}}^{}\mathcal{A}(s^t, a^{\prime}; \boldsymbol{\mathcal{W}}^t) \Bigg)
\end{align}
The primary benefit is the ability to generalize learning across actions without modifying the learning algorithm, which improves policy evaluation in the presence of numerous actions with similar state values. As a result of combining the Dueling technique and Double DQL methods (represented in Fig. \ref{fig_d3ql}), we can expect that the resultant D3QL agent will outperform its predecessors.

Finally, in practice, the agent training process described in this section could be offloaded to an external computing entity such as an edge server, or a cloud server residing in the SBS. If so, delays introduced due to the imperfection of the agent-server channel should be considered.

\subsection{Agent Customization}

The first step toward exploiting D3QL to solve problem \eqref{problem2} is to define the agent's action, reward, and state space.

\subsubsection{Action Space}
We define the action space as set $\boldsymbol{\mathcal{A}} = \{a: (r, c) | r \in \{0, ..., \mathcal{R}\}, c \in \{1, ..., \mathcal{C}\}\}$, and the action of time slot $t$ is either $a^{t}: (0, c^{t})$, which indicates sensing channel $c$, or $a^{t}: ({t}^{t} > 0, c^{t})$, which denotes the transmission of a packet with length $r$ on channel $c$. The agent controls its actual throughput on channel $c$ at each time slot by adjusting its packet length.

\textcolor{black}{\textbf{Example}: In a lightweight context, UE $0$ may choose to transmit a packet of maximum-length, say $5$, on the third second in time slot $t$. In this case, the action will be  $a_{0}^{t}: (5, 2)$.}

\subsubsection{State Space}
In the case of sensing channel $c$, the observation set would be $\boldsymbol{o}$ = \{\textit{B: Busy, I: Idle}\}, whereas it would be $\boldsymbol{o}$ = \{\textit{S: Success, C: Collision}\} in the case of packet transmission. The state of the agent, as defined in \eqref{state}, is the sequence of the most recent $\mathcal{H}$ (observation, action) tuples, along with the normalized throughputs of the agent per channel. The latter term provides a broader insight into how the agent can improve its objective function.
\begin{align}\label{state}
s^{t} = \Bigg\{ & \Big\{(o^{\tau}, a^{\tau}) | \tau \in \{ t - \mathcal{H}, ..., t \} \Big\}, \notag \\
& \Big\{ (\frac{{x}_{c}^{t}}{{\chi}_{c}^{t}}) | c \in \mathbb{C} \Big\} \Bigg\}
\end{align}
As mentioned in the previous section, we are dealing with a POMDP, thus note that the state space is a subset of the total system state available to the agent.

\textcolor{black}{\textbf{Example}: In the example above, the UE $0$ choice to transmit a maximum-length packet may stem from its previous success on the second channel and its low normalized throughputs, like: $s_{0}^{t} = \Bigg\{ \Big\{ (\textit{I}, (0, 2)), \ (\textit{S}, (4, 2)), \ (\textit{S}, (4, 2)) \Big\}, \Big\{ \frac{0.5}{0.7}, \frac{0.2}{0.7} \Big\} \Bigg\}$, where $\mathcal{H} = 3$. and $\mathcal{C}=2$}

\subsubsection{Reward}
Since the agent is designated to maximize its throughput while not deteriorating the throughputs of others, the reward should be engineered to reinforce the throughput maximization but penalize the violation of fairness. This goal is satisfied in (\ref{reward}) while providing additional guidance by reinforcing channel sensing and penalizing collisions.
\begin{equation} \label{reward}
{\rho}^{t + 1} = 
\left\{\begin{array}{ll}
    {r}^{t} - \mathcal{M} \cdot \psi^{t} & \mbox{if} \; o^{t} = S\\
    - {r}^{t} - \mathcal{M} \cdot \psi^{t} & \mbox{if} \; o^{t} = C\\
    \mu & \mbox{if} \; o^{t} \in \{ {{B, I}} \}
\end{array}\right. 
\end{equation}
In this equation, $\psi^{t}$ is $\max{\Big\{({x}_{c}^{t} - {\chi}_{c}^{t}), 0\Big\}}$, $\mathcal{M} \gg 1$ is a large constant, and $\mu \ll 1$ is a small constant to incentivize channel sensing. \textcolor{black}{The value of $\mathcal{M}$ should be sufficiently high to discourage the agent from transmitting and receiving packet-transmission rewards in unfair situations. Additionally, $\mu$ must be low enough to prevent lazy behaviors while encouraging the agent to sense the channel in uncertain situations. In our simulations, we set these values to $5$ and $0.1$, respectively.}

\textcolor{black}{\textbf{Example}: Following the previous example, the reward after successful transmission of the packet equals $5 - \mathcal{M} \cdot \max{\big\{(0.2 - 0.7), 0\big\}} = 5$. In case of an unexpected parallel transmission from another UE, caused by an inadequately trained approximator or a change in context, the reward will be $-5$.
}

\subsubsection{Approximator} 
\begin{figure}[t!]\centering
\includegraphics[width=3.4in]{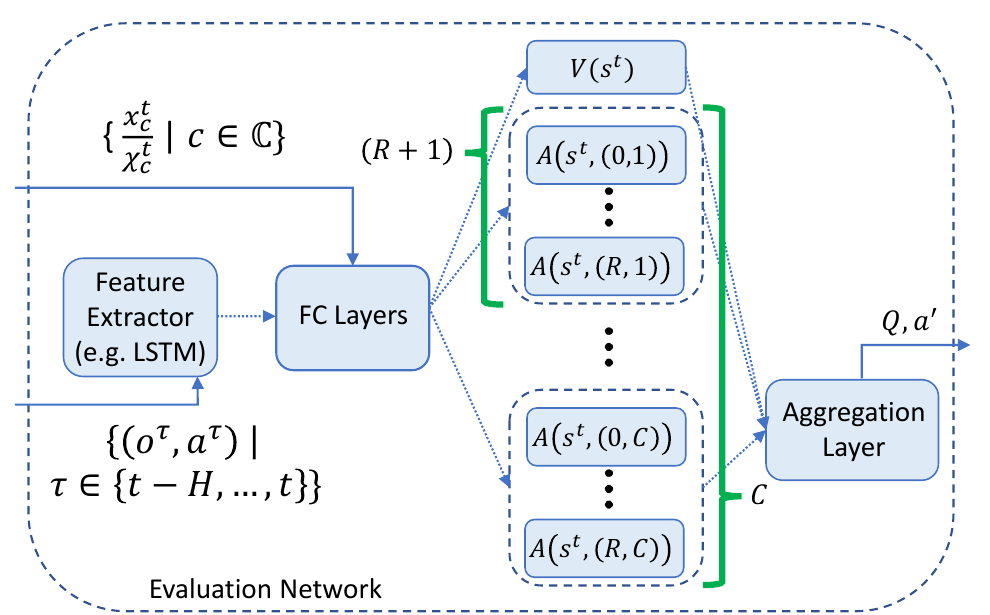}
\caption{The evaluation network of D3QL Agent (Fig. \ref{fig_d3ql}).}
\label{fig_evl_net}
\end{figure}

The evaluation network of the D3QL agent is detailed in Fig. \ref{fig_evl_net}. In this module, the first part of the state is fed to a Long Short-Term Memory (LSTM) feature extractor to discover temporal patterns. Besides, the second term in the state is directly fed to Fully Connected (FC) layers, as it provides longer-term knowledge about the system. Afterward, two sequences (or streams) of fully interconnected layers are utilized. The streams are designed to provide two separate estimators for state values and action advantages. The estimators are then combined via an aggregation layer to produce Q values. Due to the non-uniformity of actions regarding the various packet lengths, the target function in \eqref{eq_DDQL_target} must be transformed as follows:
\begin{equation}\label{eq_customized_target} 
Y_{\star}^{t}  = \frac{(1 - \gamma ^ {r^{t}})}{(1 - \gamma) \; {r^{t}}}  \;  {\rho}_{t+1} +  \gamma ^ {r^{t}}  \; \widehat{Q}(s^{t+1}, a', {\boldsymbol{\mathcal{W}}^t}^-).
\end{equation}
For actions of packet length $1$ (sensing the channel or sending a single time slot packet), \eqref{eq_DDQL_target} and \eqref{eq_customized_target} are obviously equivalent. However, future time slots are discounted for larger packages.

\textcolor{black}{\textbf{Example}: If the transition in the example above is used for training, the target value equals $\frac{(1 - 0.9 ^ {5})}{(1 - 0.9) \cdot \; {5}}  \;  {5} +  0.9 ^ {5} \cdot  \; \widehat{Q}(s^{t+1}, a', {\boldsymbol{\mathcal{W}}^t}^-)$, where $\gamma = 0.9$. Since the transmission of this packet lasts $5$ time slots, the effect of the next-state Q-value decreases.
}

\subsection{CL Mechanism}
To accommodate the non-stationary nature of the environment practically, the proposed D3QL agent should be enhanced to remember previously learned contexts and re-run the training procedure for new contexts \cite{khetarpal_towards_2022}. To accomplish this, a CL mechanism is proposed and detailed in Algorithm \ref{alg_cld3ql}. In this algorithm, $\epsilon'$ and $\widetilde{\epsilon}$ are small positive integers used to control the $\epsilon$-greedy mechanism. 
Through each step, the SBS informs the agent of any probable change in the set of active users. If so, the agent saves the current experience memory and weights before examining the recorded context references ($\boldsymbol{\Omega}$). If the current context ($\phi$) has been viewed previously (i.e. if the current context has at least one transformation that is equivalent to one of the recorded context references), the agent loads the corresponding context reference (or simply the reference, denoted by $\hat{\phi}$) along with its experience memory and weights. Otherwise, the current context will be added as a new context reference to  $\boldsymbol{\Omega}$. Next, 
the action is chosen through steps 10 to 16 following the $\epsilon$-greedy policy that follows the evaluation function of the corresponding agent with probability $(1-\epsilon)$ and chooses a random action with probability $\epsilon$. Whenever $Q$ is used, $S_{t}$ is first transformed from the state space of $\phi$ to that of $\hat{\phi}$, then the action is determined, and the selected channel is detransformed to the state space of $\phi$ so that it is applicable to the environment (since $\phi$ is currently active in the environment). Finally, the reward and observation are collected, transformed, and used to update the weights of $\hat{\phi}$ via the experience memory. During the training process, the probability decreases from $\epsilon$ to $\widetilde{\epsilon}$. 

We denote the set of active users at time slot $t$ with $\boldsymbol{\mathcal{M}}^{t} = \{ \boldsymbol{m}_{c}^{t} | c \in \mathbb{C} \}$ where $\boldsymbol{m}_{c}^{t}$ is the set of all UEs belonging to channel $c$ at time slot $t$. 
In the recognition of contexts, we exploit the permutability of $\boldsymbol{\mathcal{M}}^{t}$ over channels, where a permutation of $\boldsymbol{\mathcal{M}}^{t}$ has the same members but with different channel indices. In this way, the context defined by $\boldsymbol{\mathcal{M}}^{t}$ can be loaded and trained for all of its possible permutations. As a simple example, \textit{\{UE $0$ on channel $1$ and UE $1$ on channel $2$\}} and \textit{\{UE $0$ on channel $2$ and UE $1$ on channel $1$\}} would constitute a single context, with permutation $\{ 1 \gets 2, 2 \gets 1\}$. This approach, which is inspired by the concept of group symmetries \cite{liu2022continual}, leads to more repetitive contexts and thus, greater backward transfer capability. However, a more complex agent design is required, since the agent has to transform/detransform its actions and observations to be consistent across time.

\textcolor{black}{\textbf{Example}: If we consider the example of previous section and UE $0$ enters a new context that is the reverse of the previous one, it can easily re-use its current approximator by transforming its state into: $\hat{s}_{0}^{t} = \Bigg\{ \Big\{ (\textit{I}, (0, 1)), \ (\textit{S}, (4, 1)), \ (\textit{S}, (4, 1)) \Big\}, \Big\{ \frac{0.2}{0.7}, \frac{0.5}{0.7} \Big\} \Bigg\}$
}

\begin{algorithm}[t!]\label{alg_cld3ql}
\caption{Symmetry-aware CL-D3QL}
\KwInput{$\mathcal{T}$, $\epsilon'$, and $\widetilde{\epsilon}$}
$\boldsymbol{\Omega} \leftarrow \emptyset$, $\boldsymbol{\mathcal{W}} \leftarrow \mathbf{0}$, $\boldsymbol{\mathcal{W}^-} \leftarrow \mathbf{0}$, $\epsilon \gets 1$, $memory \gets \{\}$\\
\For{$t$ in $[0:\mathcal{T}]$}
{
    \If{new context $\phi$ is announced}
    {
        save the current context memory and weights \\
        \If{$\boldsymbol{\Omega}$ contains a transformation of $\phi$}
        {
            $\hat{\phi} \gets$ load the transformation \\
            reload $\boldsymbol{\mathcal{W}}, \boldsymbol{\mathcal{W}^-}$, and $memory$ of $\hat{\phi}$ \\
        }
        \ElseIf{$\phi \notin \boldsymbol{\Omega}$}
        {
            $\boldsymbol{\Omega} \leftarrow \boldsymbol{\Omega} \cup \{\phi\}$\\
        }
    }
    $\zeta \gets$ generate a random number from $[0:1]$ \\
    \If{$\zeta > \epsilon$}
    {
        ${\hat{s}}^{t} \gets$ transform $s^{t}$ to the state space of $\hat{\phi}$\\
        $(r, \hat{c}) \gets$ argmax$_{a \in \boldsymbol{\mathcal{A}}} Q(\hat{s}^{t}, a, \boldsymbol{\mathcal{W}})$ \\
        $c \gets$ transform $\hat{c}$ to the state space of $\phi$
    }
    \Else
    {
        select a random $(r, c)$ from $\boldsymbol{\mathcal{A}}$
    }
    transmit the packet, and get $o^{t}$ and ${\rho}^{t+1}$ \\
    calulate $s^{t+1}$ \\
    $\hat{s}^{t+1} \gets$ transform $s^{t+1}$ to the state space of $\hat{\phi}$\\
    $memory \gets memory \cup \{(\hat{s}^{t}, (r,\hat{c}), {\rho}^{t+1}, \hat{s}^{t+1})\}$ \\
    choose a batch of samples from $memory$\\
    train the agent using \eqref{eq_DQL_bellman} and \eqref{eq_customized_target}\\
    \If{$\epsilon > \widetilde{\epsilon}$}
    {
        $\epsilon \gets \epsilon - \epsilon'$
    }
}
\end{algorithm}

\subsection{Efficiency Analysis} \label{eff_analysis}
Since there is no remembering in the pure D3QL technique (without the CL mechanism), every transition introduces a new context, and the number of encountered contexts grows as the agent continues to interact with the environment. Now, if we can prove that the CL mechanism restricts the number of encountered contexts, we can expect the CL-based D3QL agent to behave more efficiently in dynamic environments by decreasing the size of its state space. Assume $\Upsilon$ is the number of unique UE types and keep in mind that $\mathcal{C}$ is the number of channels. Due to the fact that our proposed CL mechanism is symmetry-aware and UEs can be distinguished by their type (e.g., transmission profile, protocol, etc.), the number of UEs can be reduced to the number of UE types when investigating contexts.

To determine a bound for the number of contexts of the CL-D3QL mechanism, it is necessary to solve the problem of determining the number of \textit{with-replacement} selections of $\mathcal{C}$ items from $\Upsilon$ distinct items. Let's refer to the set of UE types as $\boldsymbol{\Upsilon} = \{ 1, 2, 3, ..., \Upsilon \}$. Any selection of $\mathcal{C}$ elements from the $\Upsilon$ possibilities with repetition can be described as a tuple of size $\mathcal{C}$ with non-decreasing and distinct entries, that is
\begin{equation}\label{eq_proof1} 
(\upsilon_1, \cdots, \upsilon_\mathcal{C}), \mbox{ where } 1 \leq \upsilon_1 \leq \upsilon_2 \leq \cdots \leq \upsilon_\mathcal{C} \leq \Upsilon.
\end{equation}
Now, consider that $(\eta_1, \eta_2, \cdots, \eta_\mathcal{C})$ is a new tuple of size $\mathcal{C}$ obtained from $(\upsilon_1, \upsilon_2, \cdots, \upsilon_\mathcal{C})$ as follows.
\begin{equation}\label{eq_proof2} 
(\eta_1, \cdots, \eta_\mathcal{C}) = \Big(\upsilon_1, \upsilon_2 + 1, \upsilon_3 + 2, \cdots, \upsilon_\mathcal{C} + (\mathcal{C}-1)\Big)
\end{equation}
Consequently, the subsequent conditions are met: I) $1 \leq \eta_1 < \eta_2 < \cdots < \eta_\mathcal{C} \leq \Upsilon + \mathcal{C} - 1$, II) each $\upsilon$-tuple can be represented by a unique $\eta$-tuple, and III) every tuple $(\varrho_1, \cdots, \varrho_\mathcal{C})$ of size $\mathcal{C}$ with $1 \leq \varrho_1 < \varrho_2 < \cdots < \varrho_\mathcal{C} \leq \Upsilon + \mathcal{C} - 1$ corresponds to an $\upsilon$-tuple, that is $\Big(\varrho_1, \varrho_2 - 1, \varrho_3 - 2, \cdots, \varrho_\mathcal{C} - (\mathcal{C}-1)\Big)$, which will satisfy $1 \leq \varrho_1 \leq \varrho_2 -1 \leq \varrho_3 - 2 \leq \cdots \leq \varrho_\mathcal{C} - \mathcal{C} + 1 \leq \Upsilon$. Therefore, counting $\upsilon$-tuples (that is the number of \textit{with-replacement} selection of $\mathcal{C}$ items from $\Upsilon$ distinct items) is equivalent to counting $\eta$-tuples. The advantage of this approach is that in order to count $\eta$-tuples, it is sufficient to count the number of possible $\mathcal{C}$-tuples chosen from $\{1, 2, \cdots, \Upsilon + \mathcal{C} - 1\}$ \textit{without replacement}, which is equal to $\binom{\Upsilon + \mathcal{C} - 1}{\mathcal{C}}$. It proves that the number of contexts for the symmetry-aware CL-D3QL mechanism is limited.

\begin{figure}[!t]
\centerline{\includegraphics[width=3in]{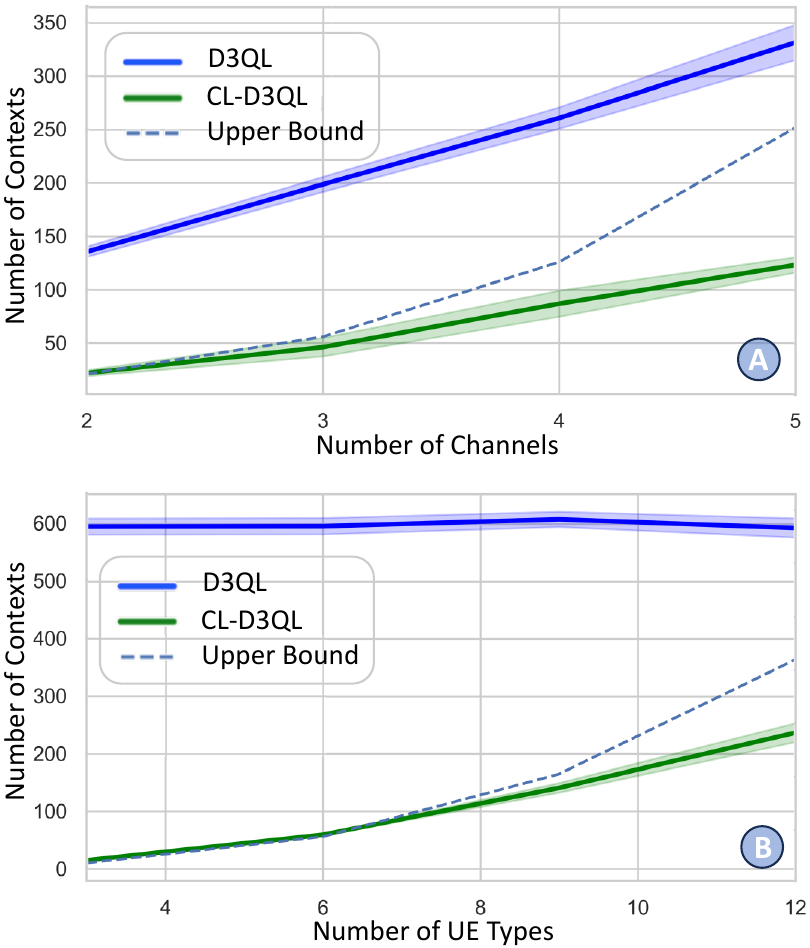}}
\caption{The maximum number of experienced contexts with the D3QL algorithm (in which any change in the set of users constitutes a new context) and the symmetry-aware CL-D3QL algorithm vs. the number of channels and the number of UE types. Note that in A, the number of UE types is $6$, whereas in B, the number of channels is $3$. In addition, the number of active UEs in both scenarios is fixed.}
\label{contexts}
\end{figure}

Fig. \ref{contexts} depicts the number of unique contexts for various numbers of channels and UE types. In this figure, the number of active UEs is fixed, and the results are captured after extremely lengthy runs. As can be seen, the calculated bound serves as the upper limit for CL-D3QL in both cases. Note that in Fig. \ref{contexts}-B, D3QL is unaware of symmetry and, regardless of the UE type, it considers each UE to be unique. Therefore, it is dependent on the (fixed) number of active UEs in the environment and not the number of UE types. It means that when the numbers of channels and active UEs do not change if the number of UE types is high, the likelihood of symmetry detection is low, and CL-D3QL would be approximately as efficient as D3QL. 

\section{Evaluation}\label{s_sim}
Within this section, a numerical analysis is conducted to investigate the effectiveness of the proposed CL-D3QL method. The hyper-parameters and configurations obtained from best practices in the literature, followed by extensive trial and error and heuristics, are listed in Table \ref{tab_sim_par}. To test the efficacy of our strategy, we carried out a series of experiments on a computer running a 64-bit operating system equipped with 16 NVIDIA Tesla V100 GPUs and 10 gigabytes of Non-Volatile Memory express (NVMe) storage. PyTorch library was utilized to effectively implement both the evaluation and target LSTM-based networks. In each experiment, comparisons are made between the CL-D3QL, D3QL, and Random algorithms. The difference between the CL-D3QL and D3QL agents is that the CL-D3QL agent has a CL mechanism, whereas the D3QL algorithm lacks remembrance, so each announced context always appears to be new to it. \textcolor{black}{Moreover, the D3QL algorithm can be considered an enhanced form of the conventional DQL algorithms, including the DLMA protocol groups (i.e., DLMA \cite{yu_deep-reinforcement_2019}, CS-DLMA \cite{yu_non-uniform_2021}, and MC-DLMA \cite{ye2021multi}) discussed in the literature review section. Remarkably, none of the mentioned algorithms are equipped with dueling and double mechanisms, nor do they support normalized throughputs for the agent in the state space to inform it about fair behavior.} Finally, the Random agent selects a random action over a random channel. This will be accomplished without any prior knowledge or any specific adjustments being made to the configuration.

To compare algorithms, we use three partially conflicting metrics: the agent's normalized throughput, collision rate, and Jain fairness index. The agent's normalized throughput is computed by summing the length of the packets successfully transmitted over the last 1000 time slots (excluding headers) and dividing it by the sum of target throughputs on all channels. The collision rate is the ratio of collision observations to total observations in the last 1000 time slots. Finally, the Jain fairness index measures the fairness among the normalized throughputs of UEs at time slot $t$ by the following formula. Note that the normalized throughputs of UEs are calculated in the same manner as the agent.
\begin{equation}\label{Jain} 
\mathcal{J}^{t} = \dfrac{\left( \sum\nolimits_{i \in \mathbb{N}} \dfrac{{x}_{i, c}^{t}}{{\chi}_{i, c}^{t}} \right)^{2}}{\mathcal{N} \cdot \sum\nolimits_{i \in \mathbb{N}} \left( \dfrac{{x}_{i, c}^{t}}{{\chi}_{i, c}^{t}} \right)^{2}}
\end{equation}
In the first scenario, we establish fixed context transition points to illustrate the efficacy of our strategy better. Then, in the second scenario, we evaluate our scheme in a more realistic and Metaverse-esque setting by assuming stochastic transition points and context specifications. \textcolor{black}{Finally, we measure the time complexity of different approaches for a variable-length scenario, along with reward plots from training, to gain insights into convergence.}

\begin{table}[t!]
\caption{Training Configuration.}
\begin{center}
\begin{tabular}{|c|c|}
\hline
\textbf{Parameter} & \textbf{Value} \\
\hline
Maximum packet length ($\mathcal{R}$) & $5$ time slots \\
Packet header size & $0.5$ time slot \\
State size ($\mathcal{H}$) & $4$ experiences \\
Capacity of experience memory & $1000$ experiences \\
Batch size & $32$ \\
Discount Factor ($\gamma$) & 0.9 \\
Learning rate & $0.001$ \\
Exploration parameters $\widetilde{\epsilon}$, $\epsilon'$ & 0.005, 0.999 \\
Approximator model & \begin{tabular}{@{}c@{}}LSTM with $64$ units + \\ fully connected with $64$ and $32$ units \end{tabular}  \\
Training frequency & \begin{tabular}{@{}c@{}} Every $10$ steps \end{tabular}  \\
\begin{tabular}{@{}c@{}} Target network update frequency \\ \end{tabular} & Every $50$ steps \\
\textcolor{black}{Reward penalties $\mathcal{M}$ and $\mu$} & \textcolor{black}{5, 0.1}  \\
\hline
\end{tabular}
\label{tab_sim_par}
\end{center}
\end{table}

\subsection{Scenario 1: Fixed Change Points}
In this scenario, it is assumed that context transitions occur at specific times, as outlined in Table \ref{tab_scr01}. TDMA($p, \tau, w$) identifies a TDMA UE that transmits a packet of length $p$ beginning on the $\tau$-th time slot of each frame with duration $w$. CSMA($p, w, w_{max}$) also identifies a CSMA UE that transmits a packet of length $p$ whenever its window size reaches $0$. The default window size and maximum window size are specified with $w$ and $w_{max}$, respectively. Lastly, CH($p, d$) relates to a channel-hopping UE with packets of length $p$ and the hopping direction of $d \in \{ 1, -1 \}$.

\textcolor{black}{This scenario represents a realization of the holographic meeting room example outlined in section \ref{vision_of_6g}, where a manager discusses a product launch using a head-mounted VR device. The device employs either CL-D3QL or D3QL schemes to adjust its transmissions. Simultaneously, a digital twin of the production line environment is supported by TDMA nodes. Additionally, to facilitate semantic-aware knowledge distribution among users, known as Knowledge Bases (KBs) \cite{shokrnezhad2023semantic}, delay-tolerant CSMA nodes are also present. Frequent transitions occur in the environment due to changes in the production line state and required KB distribution. For example, during the brainstorming phase in the third quarter, there is no need to connect to the product line, thus disabling the TDMA nodes to save resources. The following paragraph will show that using CL-D3QL instead of D3QL improves QoS/QoE for the manager during known transitions, without compromising essential TDMA and CSMA functionalities.}

\begin{table}[t!]
\caption{Scenario 1: UE Profiles and their channels on each period}
\begin{center}
\begin{tabular}{|c|c|c|c|c|}
\hline
\textbf{Profile} & \scriptsize{[0: $\mathcal{T}$/4]} & \scriptsize{[$\mathcal{T}$/4: $\mathcal{T}$/2]} & \scriptsize{[$\mathcal{T}$/2: 3$\mathcal{T}$/4]} & \scriptsize{[3$\mathcal{T}$/4: $\mathcal{T}$]} \\
\hline
TDMA(3, 0, 8) & 1 & 1 & - & 1 \\
TDMA(3, 4, 8) & 2 & 3 & - & 2 \\
CSMA(2, 4, 6) & 1 & 1 & - & 1 \\
CSMA(3, 4, 8) & 3 & 2 & 2 & 3 \\
CSMA(1, 4, 6)& 3 & 2 & 2 & 3 \\
CH(2, 1)& - & - & 1,2,3 & - \\

\hline
\end{tabular}
\label{tab_scr01}
\end{center}
\end{table}

\begin{figure*}[!t]
\centerline{\includegraphics[width=5.0in]{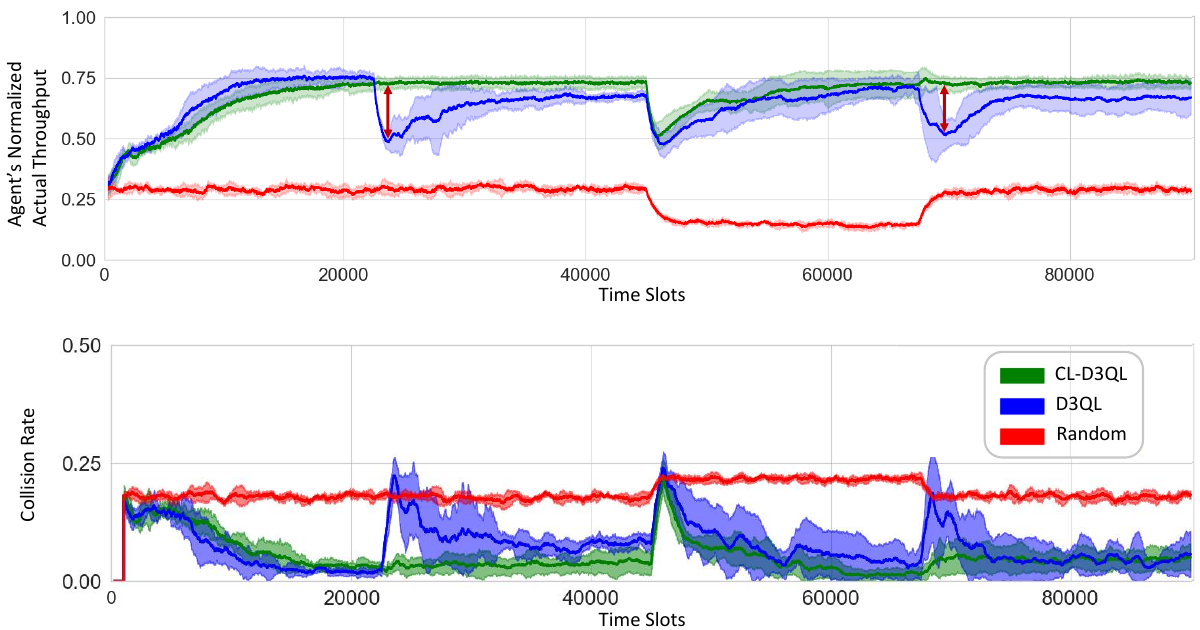}}
\caption{The agent's normalized throughput and collision rate vs. time slots for the CL-D3QL, D3QL, and Random methods in the first scenario with fixed transition points.}
\label{fig_snr01}
\end{figure*}

Obviously, the first and final quarters of the simulation take place in the same context; therefore, the CL-D3QL agent should utilize its prior knowledge of the first context when encountering it again. Perhaps less obviously, the first and second quarters of the simulation also belong to the same context, as they are permutations of each other. In other words, the agent can use the knowledge acquired during the first context to handle the second context. Fig. \ref{fig_snr01} verifies that the CL-D3QL agent possesses the required backward transfer capability for non-stationary environments, which will be extremely advantageous in controlling their dynamic nature. In addition, the figures reveal that D3QL has higher variations than CL-D3QL in all metrics, which is highly undesirable in wireless networks. Evidently, Random, the method with the lowest complexity, is also inefficient. Nonetheless, the full potential of our method would be harnessed in long-lasting scenarios, which we will demonstrate in the next section.

\subsection{Scenario 2: Stochastic Change Points}
In this scenario, context shifts occur intermittently. When a UE arrives on a channel, it remains active at a rate of $1 / \beta$ according to an exponential distribution. After its departure, a new UE will replace it, whose profile will be selected from a set of predefined profiles $\boldsymbol{\mathcal{P}}$. Three experiments are defined by hyper-parameters $\mathcal{C}$ (i.e., the number of channels), $\beta$ (i.e., the mean duration of UE existence in the network), and $|\boldsymbol{\mathcal{P}}|$ (i.e., the number of UE types), to investigate the effects of problem size, non-stationarity, and heterogeneity on the performance of our method.
\begin{itemize}
    \item \textbf{\small{Variable Number of Channels}}: As Fig. \ref{fig_scr02}-A demonstrates, the CL-D3QL agent outperforms the D3QL agent in all metrics (i.e., average normalized throughputs, collision rates, and fairness during the lifetime of the agent). Apparently, the higher fairness of random transmissions is accompanied by higher collision rates of all UEs. Moreover, with an increase in the number of channels, and hence the problem dimensions, the CL-D3QL agent perseveres and even slightly improves its performance. Scrutinizing the results revealed that this is due to the higher degree of freedom for transmissions in networks with more channels, considering that the CL-D3QL agent has more time to learn and exploit each context.
    \item \textbf{\small{Variable Context Transition Rate}}: As Fig. \ref{fig_scr02}-B illustrates, the more frequent the context transitions (lower values for $\beta$), the more continual learning improves the performance. This is due to the increased likelihood of encountering repetitive contexts, which enables the CL-D3QL algorithm to respond instantly to changes in the environment, making it suitable for the highly dynamic and non-stationary environments of the Metaverse. Nonetheless, both algorithms perform better in environments with less dynamicity.
    \item \textbf{\small{Variable Number of UE types}}: According to Fig. \ref{fig_scr02}-C, By increasing the heterogeneity of the system via increasing the number of different UE types, the CL-D3QL algorithm eventually loses its advantage over the D3QL algorithm. However, as analyzed in the previous section, it is expected that in an open-ended environment, continual learning would still outperform conventional DRL techniques.
\end{itemize}
In addition, it is important to note that, even though Random is the best-performing method in terms of fairness because it divides the spectrum equally among all nodes, it exhibits high collision rates and low normalized throughputs.

\subsection{\textcolor{black}{Time Complexity and Convergence Analysis}}

\begin{figure}[!t]
\centerline{\includegraphics[width=3.6in]{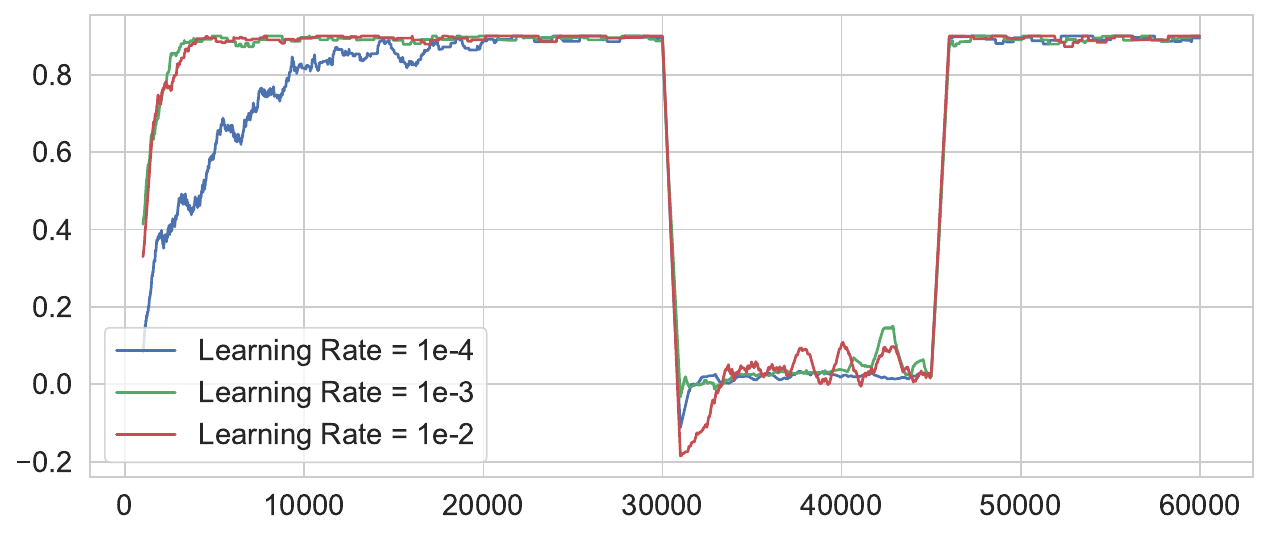}}
\caption{\textcolor{black}{Reward values for the CL-D3QL agent in the first scenario with learning rates $1e^{-3}$ (default value), $1e^{-4}$ (lower rate) and $1e^{-2}$ (higher rate).}}
\label{fig_convergence}
\end{figure}

\begin{figure*}[!t]
\centerline{\includegraphics[width=6.4in]{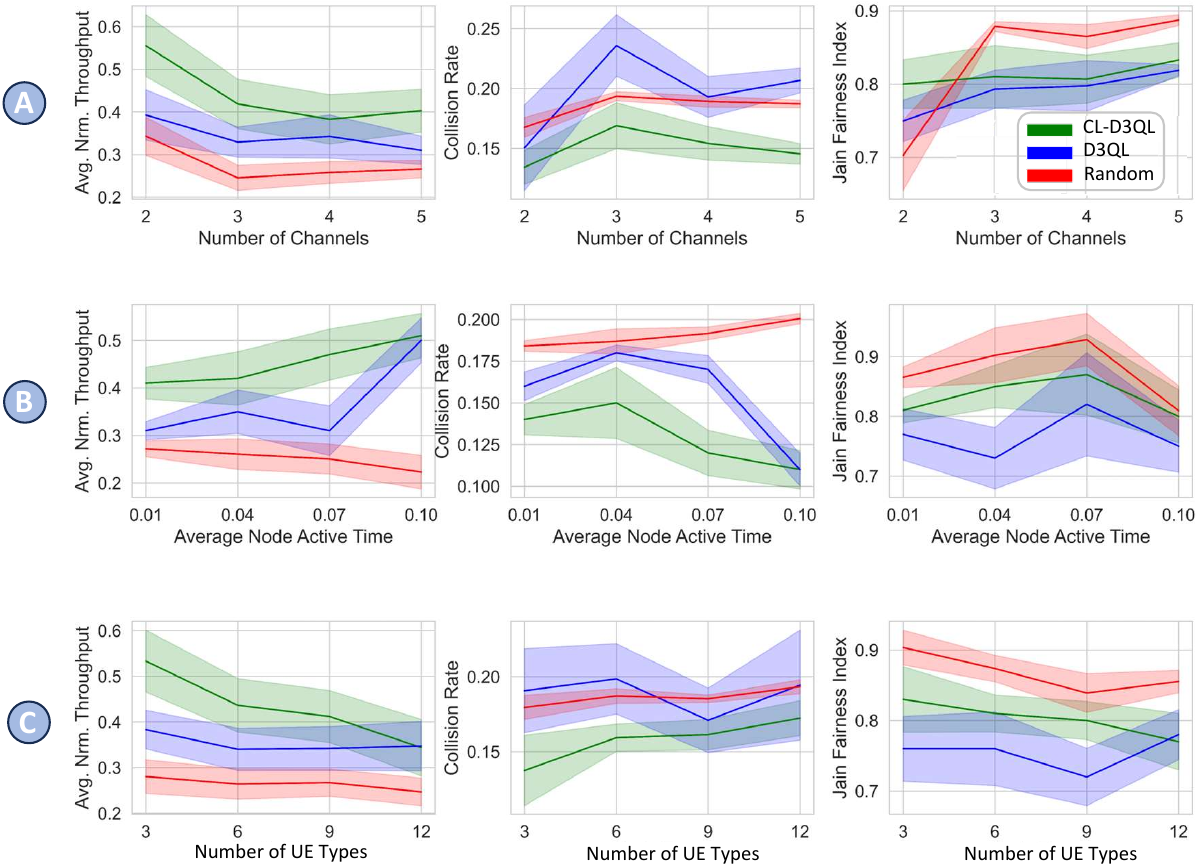}}
\caption{Average normalized throughput, collision rate, and fairness index for the CL-D3QL, D3QL, and Random agents in the stochastic scenario vs. A) the number of channels, B) the context transition rate ($1 / \beta$), and C) the number of UE Types. Each line is averaged over 5 executions, with the shaded areas being areas within the standard deviation.}
\label{fig_scr02}
\end{figure*}

\textcolor{black}{Finally, we provide a practical demonstration of our approach's time complexity, summarized in Table \ref{tab_scr02_complexity}, following the efficiency analysis in section \ref{eff_analysis}. Table \ref{tab_scr02_complexity} presents runtime data for a scenario with $\mathcal{C} =3$, $\beta = 0.04$ and $|\boldsymbol{\mathcal{P}}| = 6$ across varying durations. Notably, CL-D3QL shows execution times comparable to D3QL. In addition, the reward plots in Fig. \ref{fig_convergence} show that while the CL-D3QL algorithm converges with three learning rates of varying magnitudes, it exhibits a slow response at lower learning rates and unstable behavior during context transitions at higher learning rates. Remarkably, Fig. \ref{fig_convergence} suggests that in practical applications, a dynamic learning mechanism can be implemented where training and weight updates cease once stability is achieved for each context.}

\begin{table}[]
\begin{center}
\caption{\textcolor{black}{Time Complexity Analysis.}}
\begin{tabular}{cccc}
\hline
\textcolor{black}{Algorithm}        & \multicolumn{3}{l}{\makecell{\textcolor{black}{Approximate Time (Seconds)}}} \\
                                   & \textcolor{black}{$\mathcal{T} = 30K$}  & \textcolor{black}{$\mathcal{T} = 60K$} & \textcolor{black}{$\mathcal{T} = 90K$}   \\ \hline
\makecell{\color{black} CL-D3QL}    & \textcolor{black}{252}        & \textcolor{black}{565}    & \textcolor{black}{960}     \\
\makecell{\color{black} D3QL}       & \textcolor{black}{240}        & \textcolor{black}{541}    & \textcolor{black}{943}     \\
\makecell{\color{black} Random}     & \textcolor{black}{32 }        & \textcolor{black}{57 }    & \textcolor{black}{75 }     \\ \hline
\end{tabular}
\label{tab_scr02_complexity}
\end{center}
\end{table}

\section{Conclusion}\label{s_con}
This paper investigated the variable packet length multi-channel multiple access problem while considering a set of heterogeneous and non-stationary scenarios in which the number of active UEs and their transmission profiles might shift over time. The primary objective was to maximize the intelligent agent throughput while maintaining the performance of incumbent UEs over which the agent has no observation and agency due to backward compatibility and privacy preservation. Initially, we formulated the problem and investigated its complexity. Then, we introduced a D3QL-based agent empowered by a symmetry-aware mechanism based on DRL and CL as two Adaptive AI mechanisms that could assist in the realization of self-sustaining networks. This agent is responsible for making spectrum access decisions, such as assigning a channel and changing the packet length that must be transmitted. Afterwards, the efficiency of the CL mechanism was examined and it proved that it restricts the size of the state space and, as a result, behaves more effectively in dynamic environments. The numerical results confirmed the efficiency and fairness of the proposed agent.

As a potential future work, we intend to improve the CL-enabled D3QL-based method for accessing the spectrum for semantically-aware scenarios in which applying semantic knowledge of the environment (including UEs, resources, etc.) could help to construct parallel near-real-world experiences, which could be a game-changer for bringing the Metaverse into existence by filtering out redundant data and maximizing the use of scarce communication resources. \textcolor{black}{Furthermore, it may be worthwhile to explore practical constraints such as computational complexities and delays caused by the learning mechanism —whether on the UE side or through offloading to external servers— and their impact on the scalability of the approach. Such considerations, along with optimizing control protocols (e.g., through semantic-aware protocol learning \cite{park2024towards}), facilitate the implementation of the current scheme in a real testbed. At the same time, collaborating with industry partners allows access to real-world data and infrastructure for performance assessment under practical conditions. Finally, our future work will focus on integrating the MAC scheme with upper-layer problems, specifically predictive service provision in edge environments \cite{farhoudi2023qos} through task offloading, while accounting for task-level metrics such as task execution success rate or response delay.}

\section*{Acknowledgment}
The research work presented in this article was conducted in part at ICTFICIAL Oy. This work is partially supported by the European Union’s HE research and innovation program HORIZON-JUSNS-2023 under the 6G-Path project (Grant No. 101139172), and the European Union’s Horizon 2020 Research and Innovation Program through the aerOS project (Grant No. 101069732). The paper reflects only the authors’ views, and the European Commission bears no responsibility for any utilization of the information contained herein.

\bibliographystyle{IEEEtran}
\bibliography{IEEEabrv,main}

\begin{thebibliography}{10}
\providecommand{\url}[1]{#1}
\csname url@samestyle\endcsname
\providecommand{\newblock}{\relax}
\providecommand{\bibinfo}[2]{#2}
\providecommand{\BIBentrySTDinterwordspacing}{\spaceskip=0pt\relax}
\providecommand{\BIBentryALTinterwordstretchfactor}{4}
\providecommand{\BIBentryALTinterwordspacing}{\spaceskip=\fontdimen2\font plus
\BIBentryALTinterwordstretchfactor\fontdimen3\font minus \fontdimen4\font\relax}
\providecommand{\BIBforeignlanguage}[2]{{%
\expandafter\ifx\csname l@#1\endcsname\relax
\typeout{** WARNING: IEEEtran.bst: No hyphenation pattern has been}%
\typeout{** loaded for the language `#1'. Using the pattern for}%
\typeout{** the default language instead.}%
\else
\language=\csname l@#1\endcsname
\fi
#2}}
\providecommand{\BIBdecl}{\relax}
\BIBdecl

\bibitem{tang_roadmap_2022}
F.~Tang, X.~Chen, M.~Zhao, and N.~Kato, ``The {Roadmap} of {Communication} and {Networking} in {6G} for the {Metaverse},'' \emph{IEEE Wireless Communications}, pp. 72--81, 2022.

\bibitem{theodoropoulos2022cloud}
T.~Theodoropoulos, A.~Makris, A.~Boudi, T.~Taleb, U.~Herzog, L.~Rosa, L.~Cordeiro, K.~Tserpes, E.~Spatafora, A.~Romussi \emph{et~al.}, ``Cloud-based {XR} {Services}: A {Survey} on {Relevant} {Challenges} and {Enabling} {Technologies},'' \emph{Journal of Networking and Network Applications}, vol.~2, no.~1, pp. 1--22, Feb. 2022.

\bibitem{taleb2022vr}
T.~Taleb, N.~Sehad, Z.~Nadir, and J.~Song, ``{VR}-based {Immersive} {Service} {Management} in {B5G} {Mobile} {Systems}: A {UAV} {Command} and {Control} {Use} {Case},'' \emph{IEEE Internet of Things Journal}, vol.~10, no.~6, pp. 5349--5363, Mar. 2022.

\bibitem{taleb2022toward}
T.~Taleb, A.~Boudi, L.~Rosa, L.~Cordeiro, T.~Theodoropoulos, K.~Tserpes, P.~Dazzi, A.~I. Protopsaltis, and R.~Li, ``{Toward} {Supporting} {XR} {Services}: {Architecture} and {Enablers},'' \emph{IEEE Internet of Things Journal}, vol.~10, no.~4, pp. 3567--3586, 2022.

\bibitem{10304077}
X.~Zhou, C.~Liu, and J.~Zhao, ``{Resource} {Allocation} of {Federated} {Learning} for the {Metaverse} with {Mobile} {Augmented} {Reality},'' \emph{IEEE Transactions on Wireless Communications}, pp. 1--1, 2023.

\bibitem{10017413}
S.~K. Jagatheesaperumal and M.~Rahouti, ``{Building Digital Twins of Cyber Physical Systems with Metaverse for Industry 5.0 and Beyond},'' \emph{IT Professional}, vol.~24, no.~6, pp. 34--40, 2022.

\bibitem{Metaverse_survey_2023}
M.~Ali, F.~Naeem, G.~Kaddoum, and E.~Hossain, ``Metaverse communications, networking, security, and applications: Research issues, state-of-the-art, and future directions,'' \emph{IEEE Communications Surveys and Tutorials}, pp. 1--1, 2023.

\bibitem{giordani_toward_2020}
M.~Giordani, M.~Polese, M.~Mezzavilla, S.~Rangan, and M.~Zorzi, ``Toward {6G} {Networks}: {Use} {Cases} and {Technologies},'' \emph{IEEE Communications Magazine}, vol.~58, no.~3, pp. 55--61, 2020.

\bibitem{shokrnezhad2022near}
M.~Shokrnezhad and T.~Taleb, ``Near-optimal {Cloud}-{Network} {Integrated} {Resource} {Allocation} for {Latency}-{Sensitive} {B5G},'' in \emph{2022 {IEEE} {Global} {Communications} {Conference (GLOBECOM)}}, Rio de Janeiro, Brazil, Dec. 2022, pp. 4498--4503.

\bibitem{shokrnezhad2023scalable}
M.~Shokrnezhad, S.~Khorsandi, and T.~Taleb, ``A {Scalable} {Communication} {Model} to {Realize} {Integrated} {Access} and {Backhaul} ({IAB}) in {5G},'' in \emph{2023 {IEEE} {International} {Conference} on {Communications} ({ICC}): {Wireless} {Communications} {Symposium}}, Rome, Italy, Jun. 2023.

\bibitem{alwis_survey_2021}
C.~D. Alwis, A.~Kalla, Q.-V. Pham, P.~Kumar, K.~Dev, W.-J. Hwang, and M.~Liyanage, ``Survey on {6G} {Frontiers}: {Trends}, {Applications}, {Requirements}, {Technologies} and {Future} {Research},'' \emph{IEEE Open Journal of the Communications Society}, vol.~2, pp. 836--886, 2021.

\bibitem{abel2024definition}
D.~Abel, A.~Barreto, B.~Van~Roy, D.~Precup, H.~P. van Hasselt, and S.~Singh, ``{A definition of continual reinforcement learning},'' \emph{Advances in Neural Information Processing Systems}, vol.~36, 2024.

\bibitem{groombridge_gartner_nodate}
\BIBentryALTinterwordspacing
D.~Groombridge, ``\BIBforeignlanguage{en}{Gartner {Top} 10 {Strategic} {Technology} {Trends} for 2023},'' Gartner, Tech. Rep. [Online]. Available: \url{https://rb.gy/vrv8o}
\BIBentrySTDinterwordspacing

\bibitem{metacom_2023}
H.~Mazandarani, M.~Shokrnezhad, T.~Taleb, and R.~Li, ``{Self}-sustaining {Multiple} {Access} with {Continual} {Deep} {Reinforcement} {Learning} for {Dynamic} {Metaverse} {Applications},'' in \emph{2023 {IEEE} {International} {Conference} on {Metaverse} {Computing}, {Networking} and {Applications} ({MetaCom})}, Kyoto, Japan, Jun. 2023.

\bibitem{xu_full_2022}
M.~Xu, W.~C. Ng, W.~Y.~B. Lim, J.~Kang, Z.~Xiong, D.~Niyato, Q.~Yang, X.~Shen, and C.~Miao, ``A {Full} {Dive} {Into} {Realizing} the {Edge}-{Enabled} {Metaverse}: {Visions}, {Enabling} {Technologies}, and {Challenges},'' \emph{IEEE Communications Surveys and Tutorials}, vol.~25, no.~1, pp. 656--700, First Quarter 2023.

\bibitem{yu2024attention}
J.~Yu, A.~Alhilal, T.~Zhou, P.~Hui, and D.~H. Tsang, ``{Attention}-based {QoE}-aware {Digital} {Twin} {Empowered} {Edge} {Computing} for {Immersive} {Virtual} {Reality},'' \emph{IEEE Transactions on Wireless Communications}, 2024.

\bibitem{shokrnezhad2023double}
M.~Shokrnezhad, T.~Taleb, and P.~Dazzi, ``Double {Deep} {Q}-{Learning}-based {Path} {Selection} and {Service} {Placement} for {Latency}-{Sensitive} {Beyond} {5G} {Applications},'' \emph{IEEE Transactions on Mobile Computing}, pp. 1--14, 2023.

\bibitem{6G_Architecture}
\BIBentryALTinterwordspacing
M.~K. Bahare, A.~Gavras, M.~Gramaglia, J.~Cosmas, X.~Li, O.~Bulakci, A.~Rahman, A.~Kostopoulos, A.~Mesodiakaki, and D.~Tsolkas, ``The {6G} {Architecture} {Landscape},'' Tech. Rep., 2023. [Online]. Available: \url{https://rb.gy/vtcf0}
\BIBentrySTDinterwordspacing

\bibitem{nadir2021immersive}
Z.~Nadir, T.~Taleb, H.~Flinck, O.~Bouachir, and M.~Bagaa, ``{Immersive} {Services} {Over} {5G} and {Beyond} {Mobile} {Systems},'' \emph{IEEE Network}, vol.~35, no.~6, pp. 299--306, Nov. 2021.

\bibitem{taleb2021extremely}
T.~Taleb, Z.~Nadir, H.~Flinck, and J.~Song, ``{Extremely} {Interactive} and {Low-latency} {Services} in {5G} and {Beyond} {Mobile} {Systems},'' \emph{IEEE Communications Standards Magazine}, vol.~5, no.~2, pp. 114--119, 2021.

\bibitem{adil2024role}
M.~Adil, H.~Abulkasim, A.~Ali, H.~Song, A.~Farouk, and Z.~Jin, ``{Role of 5G and 6G Technologies in Metaverse, Quality of Service Challenges and Future Research Directions},'' \emph{IEEE Network}, 2024.

\bibitem{shokrnezhad2023semantic}
M.~Shokrnezhad, H.~Mazandarani, and T.~Taleb, ``{Semantic} {Revolution} from {Communications} to {Orchestration} for {6G}: {Challenges}, {Enablers}, and {Research} {Directions},'' in \emph{IEEE Network}, 2024.

\bibitem{shokrnezhad2024towards}
M.~Shokrnezhad, H.~Yu, T.~Taleb, R.~Li, K.~Lee, J.~Song, and C.~Westphal, ``{Towards} a {Dynamic} {Future} with {Adaptable} {Computing} and {Network} {Convergence} ({ACNC}),'' \emph{IEEE Network}, 2024.

\bibitem{matinmikko2020spectrum}
M.~Matinmikko-Blue, S.~Yrjölä, and P.~Ahokangas, ``Spectrum {Management} in the {6G} {Era}: {The} {Role} of {Regulation} and {Spectrum} {Sharing},'' in \emph{2020 2nd {6G} {Wireless} {Summit} ({6G} {SUMMIT})}, Mar. 2020, pp. 1--5.

\bibitem{yu_deep-reinforcement_2019}
Y.~Yu, T.~Wang, and S.~C. Liew, ``Deep-{Reinforcement} {Learning} {Multiple} {Access} for {Heterogeneous} {Wireless} {Networks},'' \emph{IEEE Journal on Selected Areas in Communications}, vol.~37, no.~6, pp. 1277--1290, Jun. 2019.

\bibitem{yu_non-uniform_2021}
Y.~Yu, S.~C. Liew, and T.~Wang, ``Non-{Uniform} {Time}-{Step} {Deep} {Q}-{Network} for {Carrier}-{Sense} {Multiple} {Access} in {Heterogeneous} {Wireless} {Networks},'' \emph{IEEE Transactions on Mobile Computing}, vol.~20, no.~9, pp. 2848--2861, Sep. 2021.

\bibitem{gomes2020automatic}
A.~Gomes, D.~F. Macedo, and L.~F.~M. Vieira, ``{Automatic} {MAC} {Protocol} {Selection} in {Wireless} {Networks} based on {Reinforcement} {Learning},'' \emph{Computer Communications}, vol. 149, pp. 312--323, Jan. 2020.

\bibitem{chen2022dueling}
H.~Chen, H.~Zhao, L.~Zhou, J.~Zhang, Y.~Liu, X.~Pan, X.~Liu, J.~Wei \emph{et~al.}, ``A {Dueling} {Deep} {Recurrent} {Network} {Framework} for {Dynamic} {Multi-channel} {Access} in {Heterogeneous} {Wireless} {Networks},'' \emph{Wireless Communications and Mobile Computing}, vol. 2022, Oct. 2022.

\bibitem{ye2021multi}
X.~Ye, Y.~Yu, and L.~Fu, ``Multi-{Channel} {Opportunistic} {Access} for {Heterogeneous} {Networks} {Based} on {Deep} {Reinforcement} {Learning},'' \emph{IEEE Transactions on Wireless Communications}, vol.~21, no.~2, pp. 794--807, Feb. 2022.

\bibitem{ni2024dynamic}
Y.~Ni, D.~Abraham, M.~Issa, A.~Hern{\'a}ndez-Cano, M.~Imani, P.~Mercati, and M.~Imani, ``{Dynamic MAC Protocol for Wireless Spectrum Sharing via Hyperdimensional Self-Learning},'' \emph{IEEE Access}, 2024.

\bibitem{han2024multiple}
M.~Han, Z.~Chen, and X.~Sun, ``{Multiple Access via Curriculum Multi-Task HAPPO Based in Dynamic Heterogeneous Wireless Network},'' \emph{IEEE Internet of Things Journal}, 2024.

\bibitem{park2024towards}
J.~Park, S.-W. Ko, J.~Choi, S.-L. Kim, J.~Choi, and M.~Bennis, ``{Towards semantic MAC protocols for 6G: From protocol learning to language-oriented approaches},'' \emph{IEEE BITS the Information Theory Magazine}, 2024.

\bibitem{yu2023socially}
A.~Yu, H.~Yang, C.~Feng, Y.~Li, Y.~Zhao, M.~Cheriet, and A.~V. Vasilakos, ``{Socially-aware traffic scheduling for edge-assisted metaverse by deep reinforcement learning},'' \emph{IEEE Network}, 2023.

\bibitem{yang2021brainiot}
H.~Yang, J.~Yuan, C.~Li, G.~Zhao, Z.~Sun, Q.~Yao, B.~Bao, A.~V. Vasilakos, and J.~Zhang, ``{BrainIoT: Brain-like productive services provisioning with federated learning in industrial IoT},'' \emph{IEEE Internet of Things Journal}, vol.~9, no.~3, pp. 2014--2024, 2021.

\bibitem{alwarafy2022frontiers}
A.~Alwarafy, M.~Abdallah, B.~S. Çiftler, A.~Al-Fuqaha, and M.~Hamdi, ``The {Frontiers} of {Deep} {Reinforcement} {Learning} for {Resource} {Management} in {Future} {Wireless} {HetNets}: {Techniques}, {Challenges}, and {Research} {Directions},'' \emph{IEEE Open Journal of the Communications Society}, vol.~3, pp. 322--365, 2022.

\bibitem{jadoon2022deep}
M.~A. Jadoon, A.~Pastore, M.~Navarro, and F.~Perez-Cruz, ``Deep {Reinforcement} {Learning} for {Random} {Access} in {Machine}-{Type} {Communication},'' in \emph{2022 {IEEE} {Wireless} {Communications} and {Networking} {Conference} ({WCNC})}, Apr. 2022, pp. 2553--2558.

\bibitem{doshi2021deep}
A.~Doshi, S.~Yerramalli, L.~Ferrari, T.~Yoo, and J.~G. Andrews, ``A {Deep} {Reinforcement} {Learning} {Framework} for {Contention}-{Based} {Spectrum} {Sharing},'' \emph{IEEE Journal on Selected Areas in Communications}, vol.~39, no.~8, pp. 2526--2540, Aug. 2021.

\bibitem{guo_multi-agent_2022}
Z.~Guo, Z.~Chen, P.~Liu, J.~Luo, X.~Yang, and X.~Sun, ``Multi-{Agent} {Reinforcement} {Learning}-{Based} {Distributed} {Channel} {Access} for {Next} {Generation} {Wireless} {Networks},'' \emph{IEEE Journal on Selected Areas in Communications}, vol.~40, no.~5, pp. 1587--1599, May 2022.

\bibitem{chang2023federated}
H.-H. Chang, Y.~Song, T.~T. Doan, and L.~Liu, ``{Federated} {Multi-agent} {Deep} {Reinforcement} {Learning} ({Fed-MADRL}) for {Dynamic} {Spectrum} {Access},'' \emph{IEEE Transactions on Wireless Communications}, vol.~22, no.~8, pp. 5337--5348, 2023.

\bibitem{9241414}
Y.~Liu, X.~Wang, G.~Boudreau, A.~B. Sediq, and H.~Abou-Zeid, ``{A} {Multi}-{Dimensional} {Intelligent} {Multiple} {Access} {Technique} for {5G} {Beyond} and {6G} {Wireless} {Networks},'' \emph{IEEE Transactions on Wireless Communications}, vol.~20, no.~2, pp. 1308--1320, 2021.

\bibitem{zheng2023survey}
Z.~Zheng, S.~Jiang, R.~Feng, L.~Ge, and C.~Gu, ``\BIBforeignlanguage{en}{Survey of {Reinforcement}-{Learning}-{Based} {MAC} {Protocols} for {Wireless} {Ad} {Hoc} {Networks} with a {MAC} {Reference} {Model}},'' \emph{\BIBforeignlanguage{en}{Entropy}}, vol.~25, no.~1, p. 101, Jan. 2023.

\bibitem{abbasi2021deep}
M.~Abbasi, A.~Shahraki, M.~Jalil~Piran, and A.~Taherkordi, ``Deep {Reinforcement} {Learning} for {QoS} provisioning at the {MAC} layer: {A} {Survey},'' \emph{Engineering Applications of Artificial Intelligence}, vol. 102, p. 104234, Jun. 2021.

\bibitem{watkins_q-learning_1992}
C.~J. C.~H. Watkins and P.~Dayan, ``\BIBforeignlanguage{en}{Q-learning},'' \emph{\BIBforeignlanguage{en}{Machine Learning}}, vol.~8, no.~3, pp. 279--292, May 1992.

\bibitem{mnih_human-level_2015}
V.~Mnih, K.~Kavukcuoglu, D.~Silver, A.~A. Rusu, J.~Veness, M.~G. Bellemare, A.~Graves \emph{et~al.}, ``\BIBforeignlanguage{en}{{Human-level} {Control} through {Deep} {Reinforcement} {Learning}},'' \emph{\BIBforeignlanguage{en}{Nature}}, vol. 518, no. 7540, pp. 529--533, Feb. 2015.

\bibitem{hasselt_deep_2016}
H.~v. Hasselt, A.~Guez, and D.~Silver, ``\BIBforeignlanguage{en}{Deep {Reinforcement} {Learning} with {Double} {Q}-{Learning}},'' \emph{\BIBforeignlanguage{en}{Proceedings of the AAAI Conference on Artificial Intelligence}}, vol.~30, no.~1, Mar. 2016.

\bibitem{wang2016dueling}
Z.~Wang, T.~Schaul, M.~Hessel, H.~Hasselt, M.~Lanctot, and N.~Freitas, ``{Dueling} {Network} {Architectures} for {Deep} {Reinforcement} {Learning},'' in \emph{Proceedings of The 33rd International Conference on Machine Learning}, vol.~48, Jun. 2016, pp. 1995--2003.

\bibitem{khetarpal_towards_2022}
K.~Khetarpal, M.~Riemer, I.~Rish, and D.~Precup, ``\BIBforeignlanguage{en}{Towards {Continual} {Reinforcement} {Learning}: {A} {Review} and {Perspectives}},'' \emph{\BIBforeignlanguage{en}{Journal of Artificial Intelligence Research}}, vol.~75, pp. 1401--1476, Dec. 2022.

\bibitem{liu2022continual}
S.~Liu, M.~Xu, P.~Huang, X.~Zhang, Y.~Liu, K.~Oguchi, and D.~Zhao, ``Continual {Vision}-based {Reinforcement} {Learning} with {Group} {Symmetries},'' in \emph{Proceedings of The 7th Conference on Robot Learning}, 2023, pp. 222--240.

\bibitem{farhoudi2023qos}
M.~Farhoudi, M.~Shokrnezhad, and T.~Taleb, ``{Qos-aware service prediction and orchestration in cloud-network integrated beyond 5G},'' in \emph{GLOBECOM 2023-2023 IEEE Global Communications Conference}.\hskip 1em plus 0.5em minus 0.4em\relax IEEE, 2023, pp. 369--374.

\end{thebibliography}

\vspace{1em}
\begin{wrapfigure}{l}{0.1\textwidth}
    \centering
    \vspace{-1.2em}
    \includegraphics[width=0.1\textwidth]{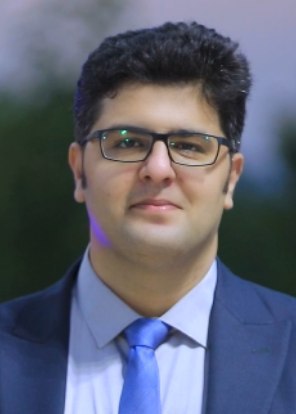}
    \vspace{-1.7em}
\end{wrapfigure}
\begin{IEEEbiographynophoto}{Hamidreza Mazandarani} (IEEE member) received his B.S. degree in telecommunications engineering from Ferdowsi University of Mashhad (FUM), Mashhad, Iran, and his M.Sc. degree (with honors) in computer systems networking from Isfahan University of Technology (IUT), Isfahan, Iran, in 2013 and 2016, respectively. Moreover, He has worked as a data scientist with various domestic and international companies. His research interests include machine learning-powered network resource allocation and semantic-aware multiple access in wireless networks.
\end{IEEEbiographynophoto}

\vspace{1em}
\begin{wrapfigure}{l}{0.1\textwidth}
    \centering
    \vspace{-1.2em}
    \includegraphics[width=0.1\textwidth]{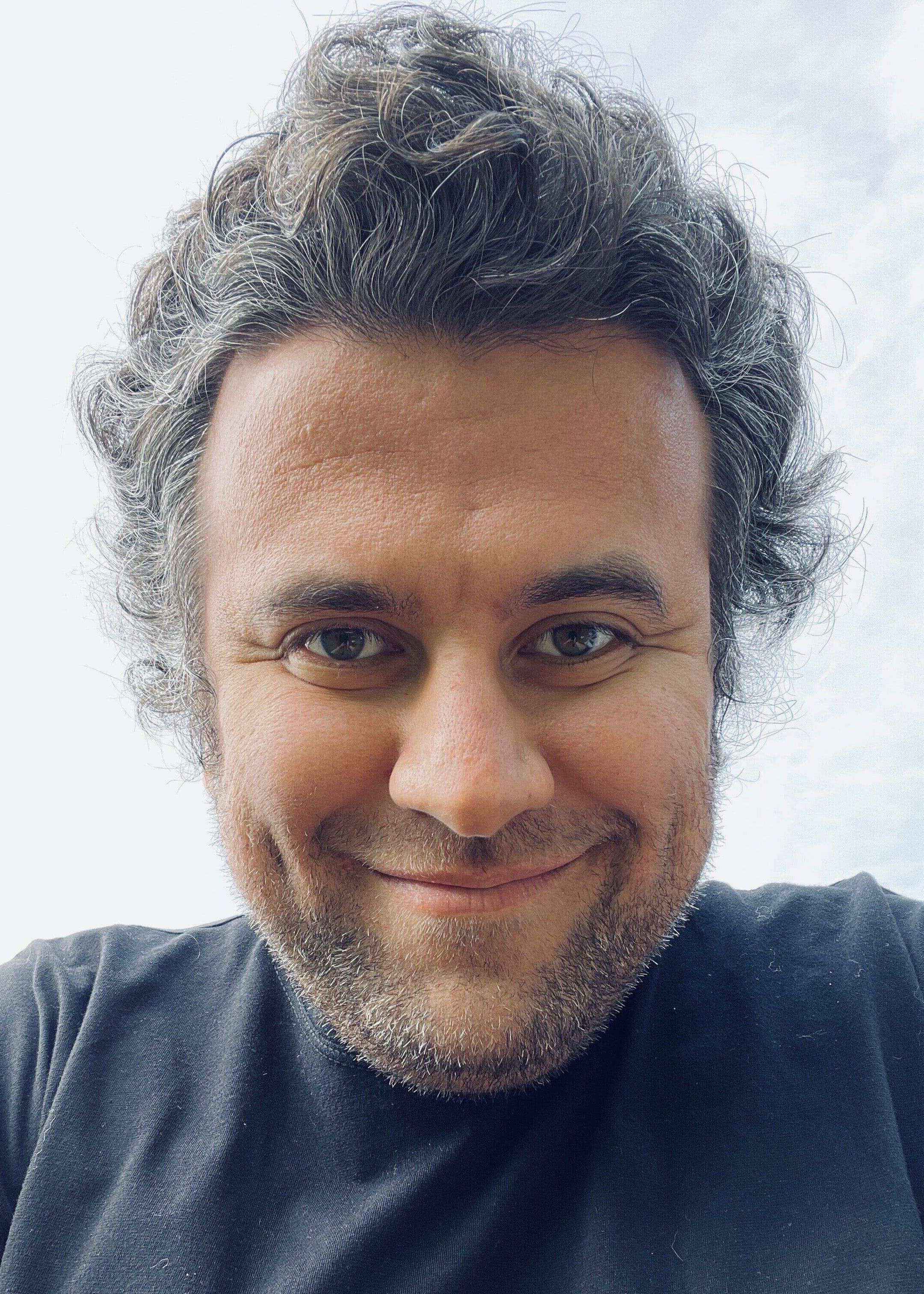}
    \vspace{-1.7em}
\end{wrapfigure}
\begin{IEEEbiographynophoto}{Masoud Shokrnezhad} received his Ph.D. degree (recognized as a bright talent) in computer networks from Amirkabir University of Technology (Tehran Polytechnic), Tehran, Iran, in 2019. He is currently a postdoctoral researcher with the Center of Wireless Communications at the University of Oulu, Finland. Between June 2021 and December 2021, he was a postdoctoral researcher at the School of Electrical Engineering, Aalto University, Espoo, Finland. Prior to that, he worked as a senior system designer and engineer with FavaPars and Pouya Cloud Technology in Tehran, Iran, since 2013. Throughout his career, Dr. Shokrnezhad has been involved in numerous national, international, and European projects focused on designing and developing computing and networking frameworks. He has also co-managed a startup that develops SDWAN solutions for B2B use cases. 
\end{IEEEbiographynophoto}

\vspace{1em}
\begin{wrapfigure}{l}{0.1\textwidth}
    \centering
    \vspace{-1.2em}
    \includegraphics[width=0.1\textwidth]{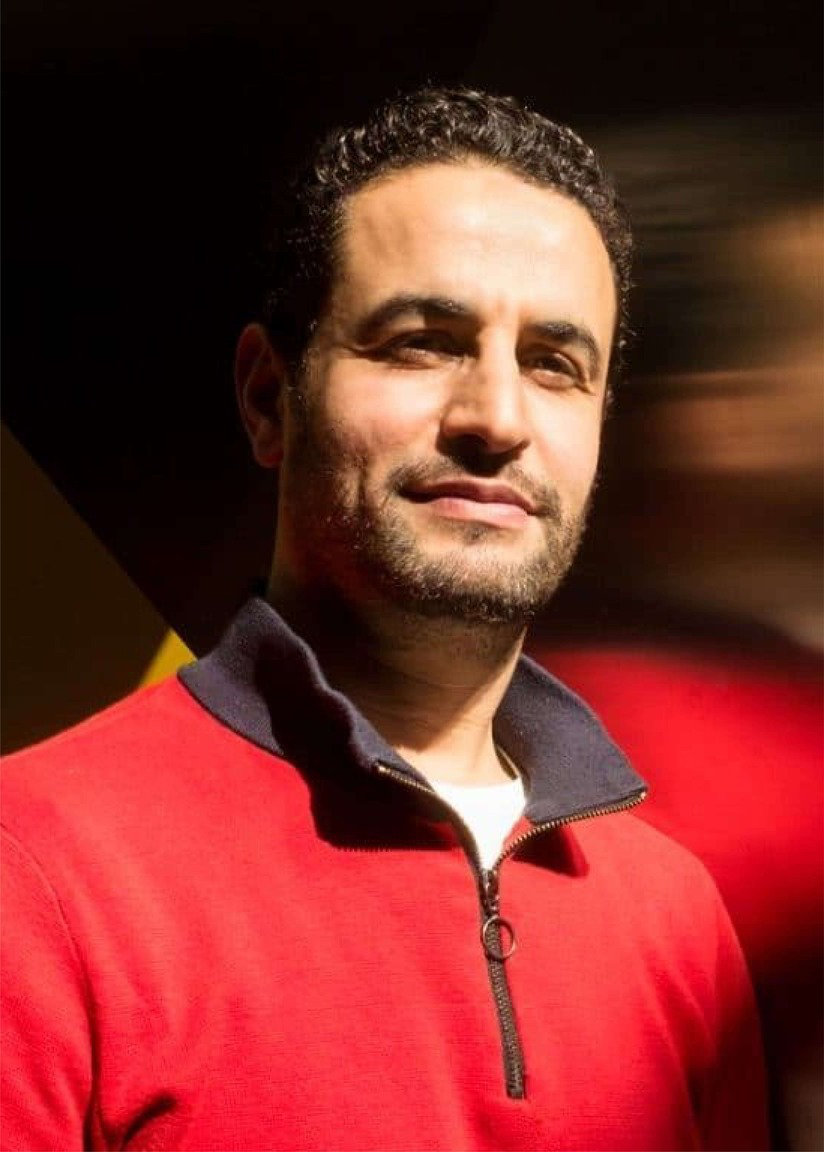}
    \vspace{-1.7em}
\end{wrapfigure}
\begin{IEEEbiographynophoto} {Tarik Taleb} is currently a Full Professor at Ruhr University Bochum, Germany. He was a Professor with the Center of Wireless Communications (CWC), University of Oulu, Oulu, Finland. He is the founder of ICTFICIAL Oy, and the founder and the Director of the MOSA!C Lab, Espoo, Finland. From October 2014 to December 2021, he was an Associate Professor at the School of Electrical Engineering, Aalto University, Espoo, Finland. Prior to that, he was working as a Senior Researcher and a 3GPP Standards Expert with NEC Europe Ltd., Heidelberg, Germany. Before joining NEC and till March 2009, he worked as an Assistant Professor with the Graduate School of Information Sciences, Tohoku University, in a lab fully funded by KDDI. From 2005 to 2006, he was a Research Fellow with the Intelligent Cosmos Research Institute, Sendai. He received the B.E. degree (with distinction) in information engineering and the M.Sc. and Ph.D. degrees in information sciences from Tohoku University, Sendai, Japan, in 2001, 2003, and 2005, respectively. 
\end{IEEEbiographynophoto}

\end{document}